\documentclass[reprint, superscriptaddress,preprintnumbers,nofootinbib, amsmath, amssymb,aps,pra,10pt]{revtex4-2}
\usepackage{graphicx}
\usepackage{dcolumn}
\usepackage{bm}
\usepackage[utf8]{inputenc}
\usepackage[DIV=13]{typearea}
\usepackage{ragged2e}
\usepackage{calligra,amsfonts,bbm,mathrsfs,amssymb,amsthm,mathtools}
\usepackage{slashed,cancel}
\usepackage{color}
\usepackage[svgnames,x11names,table]{xcolor}

\RequirePackage[ colorlinks=true,urlcolor=NavyBlue,anchorcolor=blue,citecolor=blue,filecolor=blue,               linkcolor=NavyBlue,menucolor=blue,linktocpage=true,pdfproducer=medialab,pagebackref=true]{hyperref}
\hypersetup{colorlinks=true,urlcolor=NavyBlue,anchorcolor=blue,citecolor=blue,filecolor=blue,linkcolor=NavyBlue,menucolor=blue,pdfproducer=medialab}

\usepackage[english]{babel}
\usepackage{graphicx}
\usepackage{enumerate}
\usepackage{subcaption}
\usepackage{multirow}
\usepackage[normalem]{ulem} 
\usepackage{booktabs}
\usepackage{enumitem}

\usepackage{braket}
\usepackage[skins,theorems]{tcolorbox}
\usepackage{orcidlink}
\usepackage[capitalise]{cleveref}
\usepackage[printonlyused]{acronym}
\usepackage{fdsymbol}
\newcommand{\Cwisp}{\href{https://cosmicwispers.eu/}{COSMIC WISPers} }
\tcbset{highlight math style={enhanced,
colframe=red,colback=white,arc=0pt,boxrule=1pt}}
\numberwithin{equation}{section}

\definecolor{rosso}{cmyk}{0,1,1,0.4}
\definecolor{rossos}{cmyk}{0,1,1,0.55}
\definecolor{rossoc}{cmyk}{0,1,1,0.2}
\definecolor{blu}{cmyk}{1,1,0,0.3}
\definecolor{blus}{cmyk}{1,1,0,0.6}
\definecolor{bluc}{cmyk}{1,1,0,0.1}
\definecolor{verde}{cmyk}{0.92,0,0.59,0.25}
\definecolor{verdec}{cmyk}{0.92,0,0.59,0.15}
\definecolor{verdes}{cmyk}{0.92,0,0.59,0.4}
\definecolor{pink}{rgb}{1.0, 0.75, 0.80}

\definecolor{strongcpred}{rgb}{0.8, 0.1, 0.1}
\definecolor{darkmatterblack}{rgb}{0.1, 0.1, 0.1}
\definecolor{flavourblue}{rgb}{0.1, 0.2, 0.8}
\definecolor{neutrinosgreen}{rgb}{0.0, 0.4, 0.2}
\definecolor{hierarchypurple}{rgb}{0.32, 0.12, 0.5}
\definecolor{inflationorange}{rgb}{1, 0.64, 0.}
\definecolor{darkenergy}{rgb}{0.5, 0.5, 0.5}
\definecolor{stringtheory}{rgb}{0.62, 0.12, 0.5}

\usepackage{tikz}
\usetikzlibrary{decorations.pathmorphing} 
\usepackage{halloweenmath} 
\usepackage{hyperref}

\usepackage{twemojis}
\definecolor{strongcpred}{rgb}{0.8, 0.1, 0.1}
\definecolor{darkmatterblack}{rgb}{0.1, 0.1, 0.1}
\definecolor{flavourblue}{rgb}{0.1, 0.2, 0.8}
\definecolor{neutrinosgreen}{rgb}{0.0, 0.4, 0.2}
\definecolor{hierarchypurple}{rgb}{0.32, 0.12, 0.5}
\definecolor{inflation}{rgb}{1, 0.64, 0.0} 
\definecolor{darkenergyblack}{rgb}{0, 0, 0}
\definecolor{stringtheory}{rgb}{0.62, 0.12, 0.5}
\definecolor{none}{rgb}{1, 1, 1}

\DeclareRobustCommand{\dotcolor}[1]{%
  \tikz[baseline=-0.6ex]\fill[#1] (0,0) circle (0.1cm);%
}

\DeclareRobustCommand{\squarecolor}[1]{%
  \tikz[baseline=-0.6ex]\fill[#1] (-0.1cm,-0.1cm) rectangle (0.08cm,0.08cm);%
}

\DeclareRobustCommand{\ghostcolor}[1]{%
  {\color{#1}{$\mathghost$}}%
}

\DeclareRobustCommand{\trianglecolor}[1]{%
  \tikz[baseline=-0.6ex]\fill[#1] (0,0.1cm) -- (-0.1cm,-0.1cm) -- (0.1cm,-0.1cm) -- cycle;%
}

\DeclareRobustCommand{\stringcolor}[1]{%
  \tikz[baseline=-0.6ex]%
    \draw[#1, line width=1.0pt, decorate, decoration={snake, amplitude=3.0pt, segment length=2pt}] 
      (0,0) -- (0.2,0);%
}

\newcommand{\strongcp}{%
    \texorpdfstring{\textcolor{strongcpred}{$\varheartsuit$}\,}{}
}
\newcommand{\darkmatter}{%
    \texorpdfstring{\squarecolor{darkmatterblack}\,}{DM}%
}
\newcommand{\flavour}{%
    \texorpdfstring{\textcolor{flavourblue}{$\spadesuit$}\,}{}%
}

\newcommand{\neutrinos}{%
    \texorpdfstring{\textcolor{neutrinosgreen}{$\clubsuit$}\,}{}%
}
\newcommand{\hierarchy}{%
    \texorpdfstring{\trianglecolor{hierarchypurple}\,}{Hierarchy}%
}
\newcommand{\inflation}{%
    \texorpdfstring{\textcolor{inflationorange}{$\vardiamondsuit$}\,}{}%
}
\newcommand{\darkenergy}{%
    \texorpdfstring{\ghostcolor{darkenergyblack}\,}{Dark Energy}%
}
\newcommand{\stringtheory}{%
    \texorpdfstring{\stringcolor{stringtheory}\,}{String Theory}%
}
\newcommand{\genesis}{%
    \texorpdfstring{\dotcolor{strongcpred}\,}{StrongCP}%
}

\newcommand{\none}{%
    \texorpdfstring{\squarecolor{none}\,}{DM}%
}

\newcommand{\paper}[1]{\noindent\textbf{Key Papers:} #1\par}
\newcommand{\useful}[1]{\noindent\textbf{Useful References:} #1\par}
\newcommand{\bounds}[1]{\noindent\textbf{Bounds:} #1\par}
\newcommand{\modelling}[1]{\noindent\textbf{Model:} #1\par}
\newcommand{\purpose}[1]{\noindent\textbf{Original Purpose:} #1\par}
\newcommand{\comments}[1]{\noindent\textbf{Comments:} #1\par}
\newcommand{\variation}[1]{\noindent\textbf{Variations:} #1\par}

\newcommand\subsetsim{\mathrel{%
  \ooalign{\raise0.2ex\hbox{$\subset$}\cr\hidewidth\raise-0.8ex\hbox{\scalebox{0.9}{$\sim$}}\hidewidth\cr}}}

\begin{document}
\vspace*{-10.5cm}
\hspace*{-9cm}
\preprint{
\begin{tabular*}{16.5cm}{lcc@{\extracolsep{\fill}}r}
  \vspace*{-25mm}
    \includegraphics[width=4.3cm]{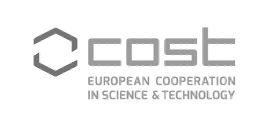}    & \includegraphics[width=4.3cm]{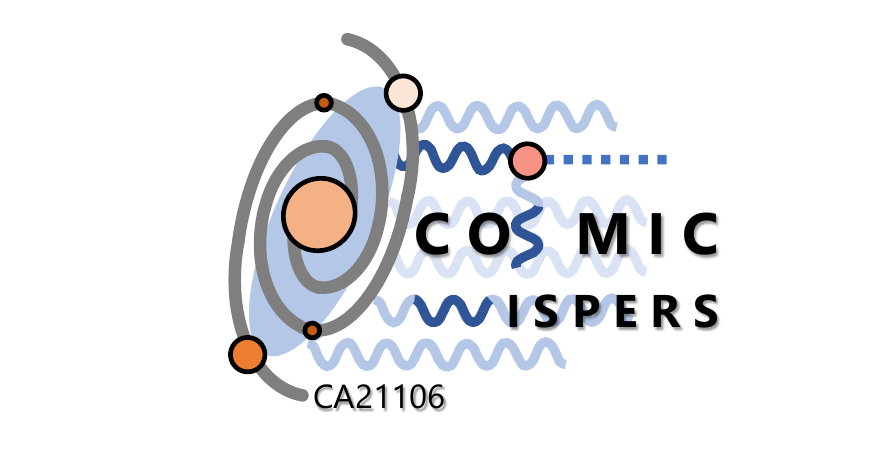}   &      \includegraphics[width=2.3cm]{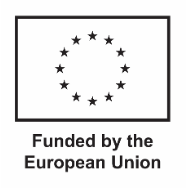} & \\
    &&& \textbf{REPORT N.}\\
    &&& IPPP/26/14\\
    &&& IFT-UAM/CSIC-26-9
\end{tabular*}
% ...
}


\title{
\vspace{1cm}
\Huge{\textbf{WISPedia}}\\
\vspace{0.3cm}
\Large{-- the WISPs Encyclopedia --}\\
\vspace{0.3cm}
\normalsize{\normalfont{Cosmic WISPers 2026 -- V1.1}}
}

\author{
Conrado Albertus \orcidlink{0000-0002-0248-8260}}
\affiliation{Department of Fundamental Physics and IUFFyM, University of Salamanca,\\ Plaza de la Merced S/N E-37008, Salamanca, Spain}

\author{
Francesca Chadha-Day \orcidlink{0000-0001-9475-8856}}
\affiliation{Institute for Particle Physics Phenomenology, Durham University, South Road, DH1 3LE, Durham, UK
}

\author{Arturo de Giorgi \orcidlink{0000-0002-9260-5466}}
\email{arturo.de-giorgi@durham.ac.uk}
\affiliation{Institute for Particle Physics Phenomenology, Durham University, South Road, DH1 3LE, Durham, UK
}

\author{
Rafid H. Dejrah \orcidlink{0000-0002-8110-296X}}
\affiliation{Department of Physics, Faculty of Science, Ankara University,  06100 Ankara, Türkiye}

\author{Marta Fuentes Zamoro~\orcidlink{0009-0001-3155-8205}}
\email{marta.zamoro@uam.es}
\affiliation{Departamento de F\'isica Te\'orica and Instituto de F\'isica Te\'orica UAM/CSIC,
Universidad Aut\'onoma de Madrid, Cantoblanco, 28049, Madrid, Spain}

\author{
Christian Käding \orcidlink{0000-0002-1781-2609}}
\affiliation{Atominstitut, Technische Universit\"at Wien,  Stadionallee 2, A-1020 Vienna, Austria}

\author{Luca Merlo~\orcidlink{0000-0002-5876-4105}}
\affiliation{Departamento de F\'isica Te\'orica and Instituto de F\'isica Te\'orica UAM/CSIC,
Universidad Aut\'onoma de Madrid, Cantoblanco, 28049, Madrid, Spain}

\author{
M. Ángeles Pérez-García \orcidlink{0000-0003-3355-3704}}
\affiliation{Department of Fundamental Physics and IUFFyM, University of Salamanca,\\ Plaza de la Merced S/N E-37008, Salamanca, Spain}

\author{Xavier Ponce D\'iaz \orcidlink{0000-0002-1305-1187}}
\email{xavier.poncediaz@unibas.ch}
\affiliation{Department of Physics, University of Basel, Klingelbergstrasse 82,  CH-4056 Basel, Switzerland}

\author{
Federico R.~Urban \orcidlink{0000-0001-9403-767X}}
\affiliation{CEICO -- FZU, Institute of Physics of the Czech Academy of Sciences, Na Slovance 1999/2, 182 00 Prague, Czech Republic}

\author{
Wen Yin \orcidlink{0000-0001-8785-6351}}
\affiliation{Department of Physics, Tokyo Metropolitan University, Tokyo 192-0397, Japan}

\begin{abstract}
The Weakly-Interacting Slim Particle encyclopedia (WISPedia) is a comprehensive reference work dedicated to the systematic compilation of theoretical models, Effective Field Theories, and frameworks involving \textit{Weakly Interacting Slim Particles} (WISPs): a broad class of light, feebly coupled particles proposed in extensions of the Standard Model. In current times, where the number of models largely surpasses the number of new physics signals, this encyclopedia aims to provide a concise reference of their landscape. The goal is to provide a useful tool to the community to navigate among them. It does \textit{not} aim to review all the models in detail, but to \textit{define} their essential characteristics, and point the reader to useful and minimal material such as the original sources, review articles, tools and general compilations of bounds. Hence, the format of this reference resembles the direct style of \textbf{a model encyclopedia of WISPs}.
\end{abstract}

\maketitle


\section*{Preface}
The search for physics beyond the \ac{SM} has led to the proposal of a vast landscape of theoretical frameworks. Among them, the family of \textit{Weakly Interacting Slim Particles} (WISPs) has emerged as a particularly rich and versatile class of candidates, capable of addressing open questions in cosmology, astrophysics and particle physics. These particles, ranging from axions and axion-like particles to hidden photons, scalars, pseudoscalars, sterile neutrinos and spin-2 particles, illustrate the growing diversity of ideas within the field. 

\begin{figure*}
    \centering
    \includegraphics[width=0.6\linewidth]{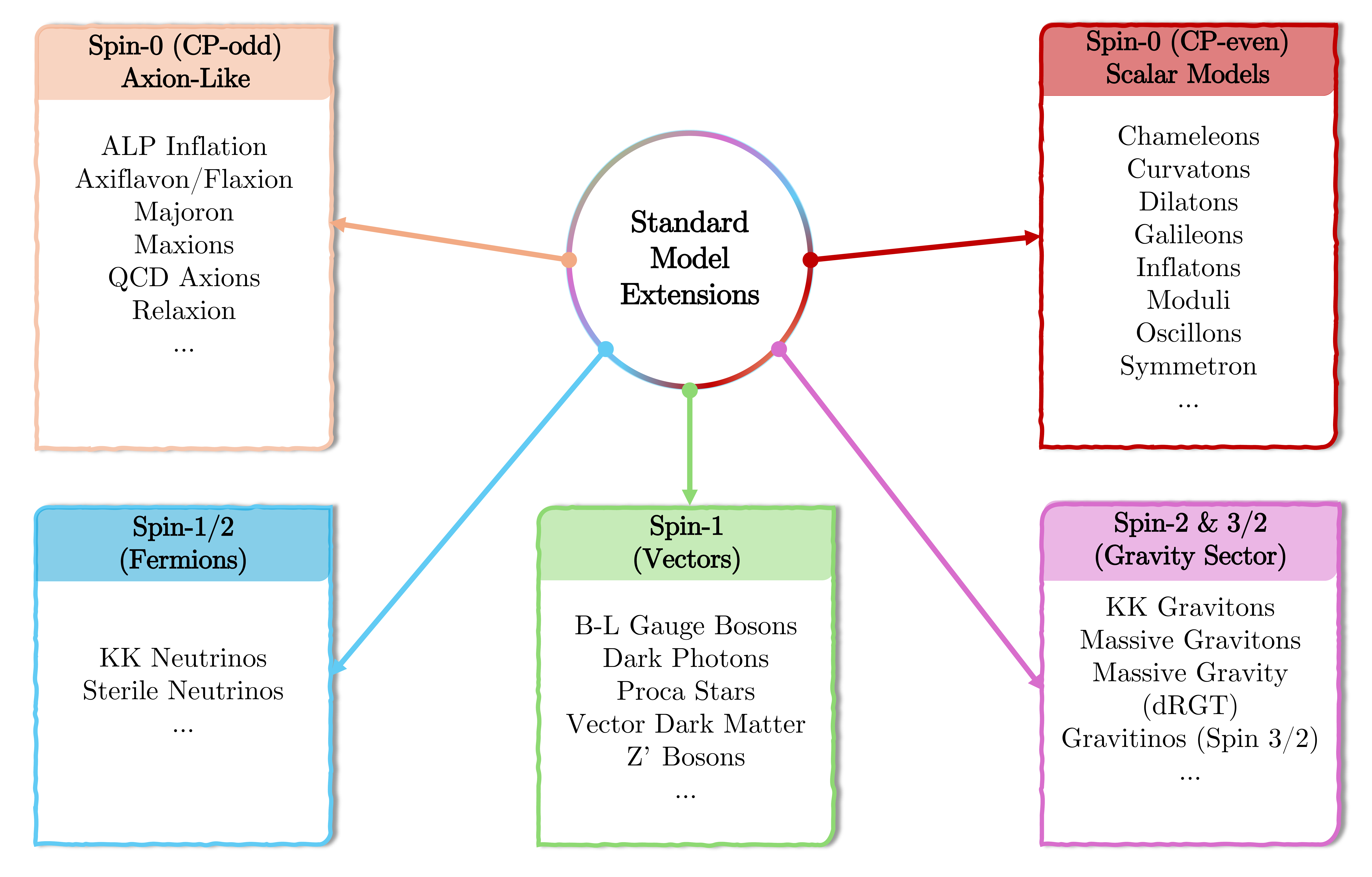}
    \caption{Schematic representation of the macro categories of the WISPedia. Note that the models listed in this figure are illustrative examples and do \textit{not} constitute an exhaustive list. For the sake of visual clarity, not every candidate discussed in the text is displayed here.}
    \label{fig:scheme}
\end{figure*}

The \textbf{WISPedia} is motivated by the need for a unified and systematic reference that organises this rapidly expanding model space. While numerous reviews exist on specific \ac{WISP} candidates or experimental searches, the goal of this work is different: to provide a concise, model-oriented encyclopedia that outlines the essential ingredients of each framework -- its particle content, interactions, and phenomenological role, while pointing the reader toward the original literature and key complementary resources. Rather than serving as an exhaustive review, the \textbf{WISPedia} aims to serve as a quick, structured gateway into the theory landscape of light, weakly coupled particles.
It also provides some information on bounds for each of them in a succinct way.

The structure of this document is intentionally modular. The macro-categories are grouped according to the spin of the new degree of freedom, as schematically depicted in Fig.~\ref{fig:scheme}. Each entry corresponds to a specific model or framework, presented in a uniform format that summarises its defining features, theoretical motivations, and relevant bounds. References to foundational papers, review articles, and useful tools are included for further exploration. The intent is to maintain clarity and accessibility while preserving the depth necessary for researchers to connect different models within the broader \ac{WISP} paradigm.\\

This encyclopedia is envisioned as a \textit{living document}, one that can evolve with the progress of the field. As new (or old but missed) models are proposed and new constraints emerge, the entries can be expanded, refined, or updated through community input. The compilation presented here has benefitted from the collective effort of the \Cwisp community, whose contributions, available in its Whitepaper~\cite{Arza:2026rsl}, have been invaluable in shaping the current content.\\

\textbf{Remarks.}
References have been cited to the best of the contributors’ knowledge. Inevitably, due to the nature of this work and its ambitious scope, some references and details may be inadvertently omitted. Furthermore, the inhomogeneous distributions among models with different spins partially reflect the interests of the \Cwisp community, rather than the global community.

Any comments, additions or corrections, particularly regarding missing references or overlooked models, are warmly welcome. The long-term success of this project relies on continuous feedback and collaboration within the \ac{WISP} research community.

\newpage

\section*{Phenomenological categories}
In order to help the reader find models of interest, we indicate with some symbols some popular applications of the model to solve open problems and puzzles of the \ac{SM}. By ``popular'' it is meant that the model is widely applied in the literature to tackle the related puzzle/problem. The lack of a symbol does \textit{not} mean that the corresponding application cannot be realised within the model.\\

The list of applications and related symbols is:
\begin{itemize}
    \item [\genesis] Baryogenesis/Leptogenesis. 
    \item [\darkenergy] \ac{DE}.
    \item [\darkmatter] \ac{DM}.
    \item [\hierarchy] \ac{EW} or Cosmological constant hierarchy structure of the \ac{SM}.
    \item [\flavour] Flavour hierarchy structure of the \ac{SM}.
    \item [\inflation] Inflation.
    \item [\neutrinos] Neutrino masses.
    \item [\stringtheory] String theory-inspired.
    \item [\strongcp] Strong CP puzzle.
\end{itemize}

At the end of the document, we present a summary with all the acronyms used throughout the paper and their definitions.

\newpage

\setcounter{footnote}{0}
\pdfbookmark[1]{Table of Contents}{tableofcontents}
\tableofcontents
\renewcommand*{\thefootnote}{\arabic{footnote}}
\newpage
\section{Spin-0}
\subsection{Axion-Like Models}
\subsubsection{ALP Anarchy\hfill\stringtheory}
\purpose{Phenomenological benchmark for the string axiverse.}
\modelling{$N$ axions with masses drawn from a logarithmic distribution and \ac{SM} couplings determined by a random $SO(N)$ rotation between the mass and coupling bases. The system with multiple \ac{ALP}s can undergo oscillations akin to neutrino oscillations.}
\paper{The framework was proposed in Ref.~\cite{Chadha-Day:2023wub}.}
\useful{Ref.~\cite{Chadha-Day:2021uyt} studies the misalignment between the \ac{ALP}s’ mass basis and the basis in which only one linear combination of \ac{ALP}s couples to electromagnetism. The framework involving multiple axions has also been studied in other contexts~\cite{Gavela:2023tzu}. More discussions about the impact of $N$ generic \ac{ALP}s in laboratory experiments such as light-shining-through-the-wall, helioscopes and haloscopes can be found in Ref.~\cite{deGiorgi:2025ldc}.}
\bounds{Bounds are modified from the single \ac{ALP} case, with bounds generally weakening as $N$ increases. Modified bounds are presented in Refs.~\cite{deGiorgi:2025ldc,Chadha-Day:2023wub, Chadha-Day:2021uyt}.}


\subsubsection{ALP EFT \hfill\none}
\purpose{The \ac{EFT} provides an effective Lagrangian of a \ac{pNGB}.}
\modelling{Agnostic approach to parametrise the \ac{ALP} interactions including all higher-dimensional operators compatible with \ac{SM} symmetries, where the \ac{ALP} is a pseudo-scalar invariant under the gauged \ac{SM} symmetries.  The Lagrangian includes shift-symmetric derivative couplings of the \ac{ALP} to fermions and anomalous ones to gauge bosons. The \ac{ALP} mass term that breaks the shift-symmetry is included \textit{ad hoc}, and it is thus a free parameter.}
\paper{The \ac{ALP} \ac{EFT} 
was proposed in Ref.~\cite{Georgi:1986df}.}
\useful{Reference~\cite{Brivio:2017ije} discusses in detail the \ac{ALP} \ac{EFT} theory at colliders' energies. The one-loop corrections to the couplings~\cite{Bonilla:2021ufe} and \ac{RGE}~\cite{Bauer:2020jbp, DasBakshi:2023lca, Chala:2020wvs,Bresciani:2024shu} have been explicitly calculated. For higher-dimensional operators, an analysis done using Hilbert series is Ref.~\cite{Grojean:2023tsd}. There are several computing tools that are useful to work with the \ac{ALP} \ac{EFT}: ALP-aca (\ac{ALP} searches in mesonic decays)~\cite{Alda:2025nsz}, ALPINIST for beamdump experiment searches~\cite{Jerhot:2022chi}, and ALPRunner (\ac{RGE})~\cite{DasBakshi:2023lca}.}
\bounds{General bounds on axions apply, e.g.\ Ref.~\cite{AxionLimits}. Additionally, further general bounds on flavour observables~\cite{Bauer:2021mvw}, mesons~\cite{diluzio:2024jip,Alda:2025uwo} and collider searches~\cite{Bauer:2017ris,Gavela:2019cmq} have been obtained.}
\comments{Two popular bases are commonly used: one where the \ac{ALP} has derivative couplings to fermions (chirality-preserving basis); and one in which it couples to fermions with Yukawa-like interactions (chirality-breaking basis). Via chiral field redefinitions of the fields, one can go from the chirality-preserving to the chirality-breaking basis, the reverse matching can only be partially performed, see Refs.~\cite{Bauer:2020jbp,Bonilla:2021ufe}. For lower energies, one has to use the Chiral Lagrangian including an \ac{ALP}~\cite{Aloni:2018vki,Bauer:2021wjo,Ovchynnikov:2025gpx,Bai:2025fvl,Balkin:2025enj}, whose latest phenomenological analysis is in Ref.~\cite{Alda:2025uwo}. The \ac{EFT} description also applies to traditional \ac{QCD} axion models. Unitarity bounds on the \ac{ALP} \ac{EFT} are collected in Ref.~\cite{Brivio:2021fog,Bresciani:2025ojh}. Explicit bottom-up constructions in which the \ac{ALP} mass is generated to match the EFT radiatively can be found in the literature, see e.g.\ Ref.~\cite{deGiorgi:2024str}.
}

\subsubsection{Anomaly-free ALP\hfill \darkmatter}
\purpose{Explaining the 3.5\,keV X-ray line anomaly~\cite{Boyarsky:2014jta,Bulbul:2014sua} with a sub-Planckian decay constant while reproducing the correct \ac{DM} abundance. It was later also applied to explain the XENON1T excess, together with the stellar-cooling hints and the \ac{DM} abundance.}
\modelling{An \ac{ALP} \ac{DM} model with a Peccei–Quinn symmetry that has no anomaly with respect to the \ac{SM} gauge group. The photon coupling is generated by a dimension-7 operator, suppressed by the mass of the lightest charged fermion that also carries the \ac{PQ} charge.}
\paper{The model was proposed and studied originally in Refs.~\cite{Nakayama:2014cza,Takahashi:2020bpq,Arias-Aragon:2020qtn}.}
\bounds{Stellar cooling, $X/\gamma$-ray observations and direct detection experiments via \ac{DM} absorption~\cite{AxionLimits,XENON:2024znc,DarkSide:2022knj,PandaX:2024cic}.}
\comments{
The excess reported by the XENON1T experiment was not confirmed by the XENONnT experiment. Several observations have also failed to detect the 3.5\,keV line. Nevertheless, for keV-scale \ac{ALP} \ac{DM} that can be searched for via absorption in Xe-type direct detection experiments, the photon coupling must be sufficiently suppressed as in the Anomaly-free \ac{ALP} to remain consistent with X/$\gamma$-ray observations~\cite{Takahashi:2020bpq,Arias-Aragon:2020qtn,Ferreira:2022egk}.
}

\subsubsection{ALP Inflation \hfill\inflation}
\label{sec:Alp_inflation}
\purpose{To realise inflation driven by an \ac{ALP} with sub-Planckian decay constants and observable couplings to \ac{SM} fields, connecting early-Universe cosmology with terrestrial experiments.}
\modelling{The inflaton is identified with an \ac{ALP} whose potential arises from multiple non-perturbative contributions. This multi-natural inflation framework achieves a sufficiently flat potential with sub-Planckian decay constants $f_\phi \ll M_{\text{Pl}}$. 
The interference between multiple non-perturbative terms can induce oscillatory features in the primordial power spectrum, serving as a distinct observational signature. Reheating occurs efficiently through direct couplings to photons $g_{\phi\gamma\gamma} \phi F\tilde{F}$ and other \ac{SM} fields, enabling experimental detection.}
\paper{The idea was proposed in Ref.~\cite{Takahashi:2019qmh}.}
\useful{See also related developments on \hyperref[sec:Alp_miracle]{\ac{ALP} miracle} inflation~\cite{Daido:2017wwb,Daido:2017tbr}, heavy \ac{QCD} axion inflation~\cite{Takahashi:2021tff}, hybrid \ac{QCD} axion inflation~\cite{Narita:2023naj} and the original multi-natural inflation concept~\cite{Czerny:2014wza,Czerny:2014xja}.}
\bounds{Predicts \ac{ALP}s with decay constants $f_\phi \sim 10^7$-$10^{13}$ GeV, masses $m_\phi \sim 0.1$-$10^6$ eV and photon couplings $g_{\phi\gamma\gamma} \sim 10^{-16}$-$10^{-10}$ GeV$^{-1}$, within reach of laboratory experiments (e.g.\ LSW, helioscopes, beam dumps) and astrophysical probes~\cite{AxionLimits}. The requirement of efficient reheating prior to \ac{BBN} ($T_{\rm reh} \gtrsim 4$ MeV) imposes a model-dependent lower bound on the coupling-mass product.}
\comments{This framework establishes the ``Big Bang on Earth" paradigm, where inflationary dynamics become testable in terrestrial experiments. 
The model naturally accommodates \ac{DM} production if the inflaton decay is incomplete or produces daughter axions, providing a unified description of inflation and \ac{DM}.
}

\subsubsection{ALP Miracle \hfill\darkmatter \inflation}
\label{sec:Alp_miracle}
\purpose{To unify inflation and \ac{DM} within a single \ac{ALP}.}
\modelling{Hilltop axion inflation, where the inflaton potential has equal magnitude but opposite curvature at the maximum and minimum. The inflaton is an \ac{ALP} that couples to photons, reheats the Universe by decaying/evaporating into plasma, while a fraction survives as \ac{DM}. The construction is agnostic, relying only on higher-dimensional operators consistent with the \ac{SM} symmetries and the upside-down symmetry.}
\paper{The model was proposed in Refs.~\cite{Daido:2017wwb,Daido:2017tbr}}
\useful{More information on hilltop inflation can be found in Refs.~\cite{Linde:1981mu,Boubekeur:2005zm,Albrecht:1982wi}. A detailed analysis of the physics potential of the International Axion Observatory (IAXO) is given by Ref.~\cite{IAXO:2019mpb}. Ref.~\cite{Takahashi:2023vhv} relates the \ac{QCD} axion with \hyperref[sec:astrophobic]{astrophobic} conditions.}
\bounds{Usual axion searches~\cite{AxionLimits} apply. Constraints from \ac{CMB} data~\cite{Planck:2018vyg, SimonsObservatory:2018koc, ACT:2025tim}, reheating and late-forming \ac{DM} are given in Ref.~\cite{Sarkar:2014bca}.}
\comments{Predicts an \ac{ALP} with mass $m_\phi \sim 0.01$-$1\,\mathrm{eV}$ and photon coupling $g_{\phi\gamma\gamma} \sim 10^{-11}\,\mathrm{GeV}^{-1}$, testable in upcoming helioscope experiments such as IAXO. It implies a small hot dark radiation component $\Delta N_{\rm eff} \sim 0.03$, potentially detectable in future \ac{CMB} observations~\cite{SimonsObservatory:2018koc, Hazumi:2019lys}.}

\subsubsection{Aximajoron/Majoraxion \hfill \neutrinos\strongcp}
\purpose{To unify the origin of neutrino masses with the solution to the strong CP problem by identifying the \ac{PQ} symmetry breaking scale with the lepton number violation scale.}
\modelling{The Peccei-Quinn scalar field acts simultaneously as the symmetry-breaking field for the seesaw mechanism, generating Majorana masses for right-handed neutrinos. Consequently, the resulting pseudo-Nambu-Goldstone boson is a hybrid of the axion and the \hyperref[sec:majoron]{Majoron}. The specific realisation depends on the embedding of the \ac{PQ} symmetry (e.g.,  \hyperref[sec:KSVZ]{KSVZ} or \hyperref[sec:DFSZ]{DFSZ} types).}
\paper{Foundational work in the context of \ac{GUTs} includes Refs.~\cite{Mohapatra:1979ia,Shafi:1984ek}. The framework has been studied for various embeddings, such as \hyperref[sec:DFSZ]{DFSZ}-type~\cite{Langacker:1986rj,He:1988dm,Geng:1988nc} and \hyperref[sec:KSVZ]{KSVZ}-type~\cite{Shin:1987xc} models.}
\useful{Recent models can be found in Refs.~\cite{Bertolini:2014aia,Clarke:2015bea,Sopov:2022bog,Arias-Aragon:2022ats}. See also \hyperref[sec:SMASH]{SMASH} model.}
\bounds{The constraints are dominated by standard axion bounds (see Ref.~\cite{AxionLimits}), as the axion-like couplings typically yield stronger limits than the neutrino sector.}
\comments{While the coupling to the neutrino sector distinguishes this particle from the standard \ac{QCD} axion, the most significant phenomenological constraints currently arise from standard axion searches (photons and nucleons).}
\subsubsection{Axion Monodromy Inflation \hfill\hierarchy}
\purpose{To achieve large-field inflation with an axion while circumventing the need for super-Planckian decay constants, generating potentially detectable gravitational waves through a monodromy mechanism that extends the field range.}
\modelling{The inflaton is an axion whose periodic potential is ``unwound" by monodromy effects in string theory, allowing the field to traverse super-Planckian distances in field space while maintaining technical naturalness protected by an underlying discrete shift symmetry. The resulting potential becomes approximately linear or power-law over large field ranges, with $p$ depending on the specific string theory realisation (D-brane, F-term or non-perturbative effects).}
\paper{The initial studies of the model include Refs.~\cite{Kallosh:2007cc, Grimm:2007hs, McAllister:2008hb}.}
\useful{A comprehensive review of axion inflation in string theory can be found in Ref.~\cite{Baumann:2014nda}, while the connection to the axiverse and multiple light fields is studied in Refs.~\cite{Arvanitaki:2009fg, March-Russell:2021zfq}. Phenomenological studies and observational predictions can be found in Refs.~\cite{Yang:2017wzh, Flauger:2009ab}.}
\bounds{It predicts an observable tensor-to-scalar ratio, (\(r\)), $r \sim 0.01-0.1$ and distinctive oscillatory features in the power spectrum and non-Gaussianity; however, the simplest realizations (e.g. linear potentials) are now in strong tension with the latest bounds from \emph{Planck} and BICEP/\emph{Keck}~\cite{Planck:2018vyg, BICEP:2021xfz} (\(r < 0.036\)) favoring models with flatter potentials (\(p< 1\)). The \ac{UV} completion in string theory imposes additional constraints on the allowed parameter space from moduli stabilisation and backreaction effects.}
\comments{This model represents a paradigmatic example of how string theory can naturally produce \ac{WISP} inflatons with protected flat potentials. The monodromy mechanism provides a technically natural explanation for large-field inflation while maintaining the axion's status as a weakly-coupled particle. Various realisations exist in string theory, including D-brane monodromy (wrapping branes on cycles with flux) and axion alignment mechanisms, each with distinct observational signatures. The model naturally connects to the broader axiverse picture, where the same axion could play multiple cosmological roles.}

\subsubsection{Astrophobic QCD Axion \hfill\darkmatter\strongcp }
\label{sec:astrophobic}
\purpose{This model suppresses axion couplings to nucleons and electrons simultaneously, the primary couplings for stellar energy loss, in order to evade stringent astrophysical bounds. This ``astrophobia'' allows the decay constant $f_a$ to be lowered to the limit imposed by the evolution of stars in the Horizontal Branch (through the photon coupling), thereby opening a parameter space accessible to terrestrial experiments.}
\modelling{Non-universal \hyperref[sec:DFSZ]{DFSZ} model, with different embeddings either with 2 or with three Higgs doublets. The \ac{PQ}-charges are chosen such that couplings to protons and neutrons are suppressed. The electron coupling must be suppressed; however, there exist several ways to implement it.}
\paper{The model is described in Refs.~\cite{DiLuzio:2017ogq,Bjorkeroth:2019jtx}. }
\useful{Due to the non-universal structure of the charges, these models exhibit flavour violation~\cite{DiLuzio:2023ndz}.}
\bounds{Usual axion searches~\cite{AxionLimits} apply; the photon coupling in particular is unaffected, hence terrestrial axion experiments can test it~\cite{Irastorza:2018dyq}. Due to the non-universal construction of these models, flavour-violating observables~\cite{MartinCamalich:2020dfe,DiLuzio:2023ndz} are important.}
\comments{Running effects were considered in Ref.~\cite{DiLuzio:2022tyc} to check that the conditions of this model are not spoiled by the running. }

\subsubsection{Clockwork QCD Axion \hfill\hierarchy\strongcp }
\purpose{Generate a \ac{QCD} axion whose interaction scale is significantly larger compared to the \ac{SSB} scale.}
\modelling{The model contains $N$ complex scalars coupled two-to-two via quartic interactions, that is, each complex scalar interacts only with the other two of them. The Lagrangian features a global $U(1)$ symmetry which acts as Peccei-Quinn, thus allowing the solution to the strong CP puzzle as in canonical \ac{QCD} models. Nevertheless, the mixing between the scalars dilutes the contribution of the \ac{QCD} potential, so that the interactions of the lightest mode, which acts as a \ac{QCD} axion, are further suppressed.}
\paper{The idea was first analysed in Refs.~\cite{Choi:2015fiu,Kaplan:2015fuy}.}
\useful{Review of the general concept of a clockwork mechanism is given in Ref.~\cite{Giudice:2016yja}.}
\bounds{Generic axion bounds applicable to \ac{ALP}s are valid, e.g.\ Ref.~\cite{AxionLimits}.}
\comments{The model generates a non-canonical \ac{QCD} axion model which deviates from the typical relation $m_a^2 f_a^2=\chi_{\text{QCD}}\approx (75.5~\text{MeV})^4$, as more broadly discussed in Ref.~\cite{Gavela:2023tzu}.}

\subsubsection*{Audible Axions \hfill \darkmatter}
\purpose{Gravitational wave production from axion/\ac{ALP} dynamics during radiation domination.} 
\modelling{The minimal scenario consists of an axion/\ac{ALP} coupled to a dark photon $X$ via the usual $ \frac{\alpha}{f_\phi} \phi F_X \tilde{F}_X$ term. The ALP is initially misaligned and starts rolling during radiation domination when its mass equals the Hubble rate, $m_\phi \approx H$. It induces a tachyonic instability in one of the dark photon helicities, resulting in exponential production of dark photons. This also amplifies quantum fluctuations of the dark photons for a range of momenta set by the \ac{ALP} mass, resulting in macroscopic anisotropic fluctuations that source gravitational waves. The resulting \ac{GW} spectrum is chiral, and different from \ac{GW}s produced e.g. in axion inflation, strongly peaked at a frequency set by the \ac{ALP} mass.} 
\paper{The idea was first studied in Ref.~\cite{Machado:2018nqk}, inspired by Refs.~\cite{Agrawal:2017eqm,Kitajima:2017peg}, where a similar setup is used to suppress the \ac{ALP} abundance.} 
\useful{A phenomenological study was performed in Ref.~\cite{Machado:2019xuc}, and a first lattice study appeared in Ref.~\cite{Ratzinger:2020oct}. Several groups included the scenario in fits to the \ac{PTA} \ac{GW} signal at low frequencies~\cite{Namba:2020kij,Ratzinger:2020koh,Madge:2023dxc,Figueroa:2023zhu,Ellis:2023oxs}.} 
\bounds{Observable \ac{GW}s are produced for large \ac{ALP} decay constants, $f_\phi \gtrsim 10^{16}~{\rm GeV}$. The full range of \ac{ALP} masses is allowed, interesting values being $m_\phi \sim 10^{-14}$~{\rm eV} for \ac{GW}s in the \ac{PTA} band~\cite{Madge:2023dxc,Figueroa:2023zhu,Ellis:2023oxs}, $m_\phi \sim 10^{-4}$~{\rm eV} for space based \ac{GW} detectors (LISA) and $m_\phi \sim$~{\rm keV} for ground based detectors.} 
\comments{The minimal scenario suffers from overproduction of \ac{ALP} \ac{DM} in some regions of parameter space. This is resolved in scenarios with rotating \ac{ALP}s~\cite{Co:2021rhi,Madge:2021abk}, where
a broader range of decay constants becomes accessible, and allows a connection with baryogenesis through axiogenesis~\cite{Co:2019wyp}.}

\subsubsection{Composite Axion \hfill\hierarchy\strongcp}
\purpose{
Solve the strong CP problem dynamically without fundamental scalars, avoiding the introduction of extra hierarchy problems.
}
\modelling{An ``axicolor" confining force with axiquarks will give rise to an axion after the dynamical breaking of the symmetry (\textit{à la} \ac{QCD} chiral symmetry breaking). These underlying strong dynamics play the dual role of spontaneously breaking the global symmetry and generating the scale of the axion decay constant. Axiquarks emerge as vector-like heavy fermions. The specific dynamics will depend on the specific non-Abelian group employed.}
\paper{The idea was first proposed in Refs.~\cite{Kim:1984pt,Choi:1985cb}.}
\useful{An additional study on the model was given in Ref.~\cite{Kaplan:1985dv}. Composite axions models can implement an accidental $U(1)_{PQ}$ symmetry by further enlarging the \ac{SM} gauge group \cite{Randall:1992ut,Redi:2016esr,Lillard:2018fdt} or by considering chiral confining dynamics such as $SU(5)$ chiral~\cite{Gavela:2018paw}, supersymmetric $SU(10)$ chiral~\cite{Gherghetta:2025kff} and Pati-Salam groups~\cite{Gherghetta:2025fip}.}
\bounds{General \ac{ALP} limits apply~\cite{AxionLimits}.}
\comments{It is important to consider the Landau pole of the new group to understand if the theory is valid.}

\subsubsection{CP-even ALP \hfill\darkmatter}
\purpose{Construct a simple renormalisable model in which an \ac{ALP} couples to the \ac{SM} without imposing CP symmetry, as a testable \ac{ALP} model at accelerators, and as the \ac{DM} candidate.}
\modelling{Introduce a new global $U(1)$ symmetry and a complex scalar whose renormalisable couplings to the \ac{SM} are allowed, which thus only allows the Higgs-portal interaction. A \ac{pNGB}, $a$, is associated with the explicit $U(1)$-breaking term, controlled by a single order parameter. Due to generic CP-violating terms in the Higgs potential, in the low-energy effective interactions, $a$ behaves as a CP-even particle due to the accidental symmetry; in particular, Higgs-portal mixing generates scalar couplings to \ac{SM} fields. }
\paper{Refs.~\cite{Sakurai:2021ipp,Haghighat:2022qyh} discuss the model in detail.}
\bounds{Bounds from colliders come from exotic Higgs decays ($h\to aa$) with prompt or displaced $a$ decays, as well as from the invisible and semi-invisible Higgs widths~\cite{Sakurai:2021ipp,Haghighat:2022qyh}. Astrophysics and cosmology constraints come from: (i) Stellar-cooling and supernova energy-loss limits on light scalars with Higgs mixing~\cite{AxionLimits}; (ii) X/$\gamma$-ray limits~\cite{AxionLimits} and small-scale-structure constraints~\cite{Essig:2013goa} when $a$ is \ac{DM} produced thermally (requiring sufficiently long lifetime and weak couplings).} 
\comments{The dominant coupling to \ac{SM} particles arises via Higgs mixing and can be extremely small for light $a$, allowing cosmologically stable \ac{DM} while keeping collider signals possible through exotic Higgs decays. 
}

\subsubsection{CP-violating Axion/ALP \hfill \darkmatter\genesis}
\purpose{Extend original ALP/axion models to include new sources of CP violation beyond the \ac{SM} which can manifest as CP-violating interactions. These interactions open new discovery windows via macroscopic forces and permanent \ac{EDMs}.}
\modelling{While the standard axion is a pseudoscalar boson, additional CP-violating sources (such as non-zero $\theta_{\rm eff}$ or \ac{UV}-originating operators) induce a scalar coupling to fermions, $g_S \bar{\psi}\psi a$. This scalar component allows the axion to mediate a spin-independent macroscopic force (monopole-monopole) or a spin-dependent force (monopole-dipole) distinct from standard axion phenomenology. In the \ac{ALP} limit, general CP-violating operators are effectively described by dimension-5 operators~\cite{DiLuzio:2023lmd}.}
\paper{The foundational proposal of macroscopic forces mediated by CP-violating axions is Ref.~\cite{Moody:1984ba}. The CP-violating axion-nucleon interactions were studied in Refs.~\cite{Pospelov:1997uv,Raffelt:2012sp,OHare:2020wah}.}
\useful{A comprehensive review of the theory for CP-violating axions and \ac{ALP}s can be found in Ref.~\cite{DiLuzio:2023lmd}, while Ref.~\cite{DiLuzio:2021jfy} presents a short summary. The modified chiral Lagrangian entailing a CP-violating \ac{ALP} was studied in Ref.~\cite{DiLuzio:2023cuk}. In Ref.~\cite{DiLuzio:2020oah}, the impact on the \ac{EDMs} of molecules, atoms, nuclei and nucleons is systematically studied up to two-loops.}
\bounds{Its coupling is stringently constrained by tests of the Equivalence Principle (torsion pendulums, E\"{o}t-Wash), short-range gravity experiments and searches for nucleon and atomic \ac{EDMs}~\cite{DiLuzio:2020oah}. See Ref.~\cite{Ramadan:2024vfc} for a recent study.}

\subsubsection{DFSZ QCD Axion \hfill\darkmatter \strongcp }
\label{sec:DFSZ}
\purpose{Solve the strong CP problem dynamically.}
\modelling{Two doublets of $SU(2)_L$ (\textit{à la} \ac{2HDM}) that couple to the up and down-quark sector respectively, and a complex singlet. Upon \ac{SSB}, the axion will arise as a mixture of the electrically neutral degrees of freedom present.}
\paper{The model was discussed in Refs.~\cite{Zhitnitsky:1980tq,Dine:1981rt}.}
\useful{A detailed discussion of the landscape of \ac{QCD} axion models beyond this one can be found in Ref.~\cite{DiLuzio:2020wdo}. Reference~\cite{GrillidiCortona:2015jxo} studies several properties of the \ac{QCD} axion at high precision, as well as the temperature dependency.}
\bounds{General \ac{ALP} limits apply~\cite{AxionLimits}.}
\comments{It can be a candidate for \ac{DM} via misalignment~\cite{Abbott:1982af,Dine:1982ah,Preskill:1982cy} (see \ref{sec:misalignment}). It is subject to the quality and \ac{DW} problems. The original model presents a family-universal charge arrangement for the \ac{SM} fermions. However, a non-universal arrangement allows to avoid the \ac{DW} problem by finding DFSZ models with $N_{\mathrm{DW}}$. The original idea can be traced back to Ref.~\cite{Geng:1988nc}; in more recent literature, all models with $N_{\mathrm{DW}}=1$ have been classified in Ref.~\cite{Cox:2023squ} and the phenomenological interplay between the heavy modes and light axion studied in Ref.~\cite{DiLuzio:2023ndz}. More generic bounds can be extracted from Refs.~\cite{MartinCamalich:2020dfe,DiLuzio:2023ndz}. This type of construction has also been used to build \hyperref[sec:astrophobic]{astrophobic} axion models, where couplings to first-generation fermions is suppressed and astrophysical bounds much reduced~\cite{DiLuzio:2017ogq,Bjorkeroth:2019jtx}.}

\subsubsection*{Early Oscillating Axions \hfill \darkmatter}\label{sec:misalignment}
\purpose{
Producing axion fields as \ac{DM} in a minimal manner.
}
\modelling{
The misalignment (or realignment) mechanism is a built-in mechanism in most light scalar \ac{DM} models, especially for axions. 
A homogeneous scalar field begins coherent oscillations once the Hubble parameter and the scalar mass become comparable. 
These coherent oscillations contribute to the \ac{DM} abundance.
}
\paper{
It was originally proposed in Refs.~\cite{Preskill:1982cy,Abbott:1982af,Dine:1982ah} in the context of the axion.
}
\useful{
In particular, for the axion, the abundance depends on the initial field value, and various scenarios for setting this initial value have been studied.
\begin{itemize}
\item \textit{Stochastic axion scenario~\cite{Graham:2018jyp,Guth:2018hsa}}: 
The inflationary equilibrium distribution of the axion is taken into account to suppress the abundance. 
This allows the \ac{QCD} axion to be the dominant \ac{DM}, for example, in certain string-theoretic realizations.
\item \textit{Hilltop axion scenario~\cite{Lyth:1991ub,Kobayashi:2013nva,Arvanitaki:2019rax}}: 
The axion is placed near the potential hilltop by another dynamical field~\cite{Daido:2017wwb,Co:2018mho,Takahashi:2019pqf}, enhancing the abundance.
\end{itemize}
}
\bounds{
If the inflation scale is too high and the scalar \ac{EFT} is already realized during inflation, an isocurvature constraint applies (see e.g. Refs.~\cite{Lyth:1991ub,Kobayashi:2013nva}). 
Moreover, for axions, the onset of oscillations must occur sufficiently early so as not to interfere with small-scale structure formation, which sets a generic constraint on the product of the mass and decay constant~\cite{Marsh:2019bjr,Dror:2020zru}.
}

\subsubsection{Early Rotating Axions \hfill \darkmatter}
\purpose{
Open a large region of parameter space for ALP dark matter that will be tested in upcoming experiments.}
\modelling{They rely on a complex scalar
field $P= S e^{\phi/f_a}$, corresponding to the \ac{PQ} field in
the case of the QCD axion, where $S$ and $\phi$ describe
the radial and angular field directions. The latter is identified
with the axion.
The key evolution is that the dynamics start
at a large $S$ field value of the \ac{PQ} field rather than at the origin of the potential. A kick is transferred to the axion from the radial mode motion. The axion can acquire large kinetic energy, which delays the time when it will eventually be trapped by its potential barriers. Axion oscillations start much later, such that the axion energy is less redshifted than in the standard misalignment mechanism {(see \ref{sec:misalignment})}, enabling to achieve the correct dark matter abundance for a much lower value of $f_a$.}
\paper{In Ref.~\cite{Eroncel:2022vjg} it was realised that the condensate fragments in most of the parameter space, leading to interesting phenomenological signatures such as axion mini-clusters~\cite{Eroncel:2022efc}. Unique gravitational-wave signatures have also been studied in  Refs.~\cite{Gouttenoire:2021wzu,Co:2021lkc,Gouttenoire:2021wzu}.}
\useful{The kinetic misalignment mechanism was first proposed in Refs.~\cite{Co:2019jts,Chang:2019tvx,Co:2020dya}. Rotating axions can also lead to another source of axion production: Axion fluctuations are generated by the product of the curvature perturbation and the fast-rolling background axion  fields~\cite{Eroncel:2025qlk,Bodas:2025eca}. {Domain wall formation was discussed in the context of axion rotation induced by level crossing with another axion~\cite{Daido:2015cba,Daido:2015bva}.}}
\bounds{Detailed model realisations and constraints have been analysed in detail in Ref.~\cite{Eroncel:2024rpe}.}
\comments{This scenario requires an explicit $U(1)$–breaking
term that operates only at early times and kicks the complex
field into an elliptic orbit as well as a radial-motion damping mechanism to bring the field orbit into a circle, which later
settles down to the potential minimum.}

\subsubsection{Discrete Goldstone Bosons \hfill\hierarchy}
\purpose{Reduce the ultraviolet
sensitivity of \ac{NGB}s masses.}
\modelling{\ac{EFT} description of scalar fields belonging to irreducible real representations of discrete symmetries $D$. The corresponding scalar potential describes a \ac{SSB} of continuous symmetries and non-linearly realises the discrete symmetries $D$. This approach leads to a hierarchy between the mass of the scalar and the \ac{UV} cutoff.}
\paper{The idea was developed and studied in Refs.~\cite{Hook:2018jle,DiLuzio:2021pxd,Vileta:2022jou}.}
\useful{More information on this model can be found in Refs.~\cite{DiLuzio:2021gos,Das:2020arz}.}
\bounds{Limits come from the degeneracy of the discrete \ac{NGB}s, as well as simultaneous productions of the discrete \ac{NGB}s at colliders~\cite{Vileta:2022jou}.}

\subsubsection{Feebly Interacting Peccei–Quinn Model \hfill \darkmatter\hierarchy\strongcp}
\purpose{Addressing the strong CP problem via the \ac{QCD} axion, while alleviating both the quality problem and the hierarchy problem.}
\modelling{From the viewpoint of the large scale required for the \ac{QCD} axion decay constant constrained by stellar-cooling observations and the quality problem, the wave function of the \ac{PQ} Higgs field is taken to be extremely large instead of assuming a large mass scale as in the conventional \ac{PQ} model. As a result, the quality problem and hierarchy problem are alleviated. The \ac{PQ} Higgs boson and \ac{PQ} fermions (in the KSVZ-type model) become light.}
\paper{The model was proposed in Ref.~\cite{Yin:2024txg}.}
\bounds{The usual bounds for the \ac{QCD} axion apply~\cite{AxionLimits}, along with those for a light \ac{PQ} Higgs boson that couples to the colored sector. 
}
\comments{The cosmology of the axion and the \ac{PQ} Higgs boson differs from that of the original model. The model construction with a large wave function needs to be explored. In this scenario, a negative portal coupling between the \ac{PQ} Higgs and usual Higgs fields naturally leads to the formation of short-lived topological defects~\cite{Yin:2024pri}, avoiding both the domain wall and isocurvature problems.}

\subsubsection{Flaxion/Axiflavon \hfill\flavour\strongcp}
\purpose{Provides a unified solution to the strong CP problem and the \ac{SM} flavour puzzle by identifying the \ac{PQ} symmetry with a \ac{FN} horizontal symmetry.}
\modelling{The \ac{PQ} symmetry $U(1)_{\text{PQ}}$ acts as a flavour-dependent \ac{FN} symmetry $U(1)_{\text{FN}}$. A complex scalar singlet $\phi$ (the flavon) breaks this symmetry at a scale $f_a$. The angular mode of $\phi$ is the \ac{QCD} axion. Yukawa couplings, forbidden at dimension-4, are generated via non-renormalisable operators involving the flavon \ac{VEV}, 
$(\langle \phi \rangle /\Lambda)^{n_{i}}$, where $\Lambda$ is the cutoff scale. The hierarchy of fermion masses and mixings is naturally explained by the non-universal charges of the fermions, which determine the power $n_{i}$.}
\paper{The simultaneous solution was independently proposed in~\cite{Ema:2016ops} (Flaxion) and~\cite{Calibbi:2016hwq} (Axiflavon). Early motivation can be traced back to Refs.~\cite{Davidson:1981zd,Wilczek:1982rv}.}
\useful{For the foundational mechanism of generating flavour hierarchies, see the original \ac{FN} paper~\cite{Froggatt:1978nt}.}
\bounds{In addition to standard axion limits (stellar cooling, SN1987A)~\cite{AxionLimits}, this model is strongly constrained by \ac{FCNCs} induced by the non-diagonal axion couplings. The golden channel is often the decay $K^+ \to \pi^+ a$~\cite{MartinCamalich:2020dfe}, with complementary bounds from $B$-physics and beam-dump experiments.}

\comments{The domain wall number $N_{\text{DW}}$ is typically non-trivial in this type of construction and often requires pre-inflationary symmetry breaking or explicit breaking mechanisms to avoid cosmological problems~\cite{Davidson:1983tp,Davidson:1984ik}. Recent work has explored discrete symmetries leading to \ac{FN}-\ac{ALP} scenarios~\cite{Greljo:2024evt}.}

\subsubsection{Fuzzy axion \hfill\darkmatter}
\purpose{Explain \ac{DM} via an ultralight axion, with wave-like behaviour due to the largeness of its de Broglie wavelength.}
\modelling{Scalar field with $10^{-24}\, \textrm{eV}\lesssim m\lesssim 10^{-18}\, \textrm{eV}$, where \(m\) represents the mass of the axion particle, minimally coupled and comprising all \ac{DM}.}
\paper{The model was proposed in Ref.~\cite{Hu:2000ke}}
\useful{For reviews on the topic, check Refs.~\cite{Eberhardt:2025caq, Niemeyer:2019aqm,Hui:2021tkt,Chadha-Day:2021szb,Marsh:2015xka}. }
\bounds{Generic axion bounds are applicable~\cite{AxionLimits}. Additional constraints applicable to \ac{ALP} ultralight \ac{DM} come from cosmic birefringence~\cite{Minami:2020odp, Nakatsuka:2022epj, Zhang:2024dmi} and superradiance~\cite{Unal:2020jiy}.}
\comments{Other candidates for fuzzy \ac{DM} have been proposed and studied in the literature (for a review, check~\cite{Eberhardt:2025caq}). The mass range of fuzzy \ac{DM} required to solve the small-scale structure anomalies ($m \sim 10^{-22}$ eV) is in strong tension with constraints from the Lyman-$\alpha$ forest, which favour $m \gtrsim 10^{-21}$ eV.}

\subsubsection{Heterotic string theory axiverse \hfill\stringtheory}
\purpose{Investigates the axiverse within the framework of the heterotic string.}
\modelling{Axions emerge from higher-form gauge fields and complex scalars. Their coupling to gauge bosons is fixed in the 10d supergravity theory by the requirement of anomaly cancellation.}
\paper{Recent studies of the expected axion masses and decay constants have been provided in~\cite{Leedom:2025mlr}. Refs.~\cite{Agrawal:2024ejr,Reig:2025dqb} study their couplings to the SM gauge bosons as a way to test this kind of string theory. }
\useful{See Ref.~\cite{Svrcek:2006yi} for a detailed study of axions in different heterotic strings, both in weak and strong coupling. }
\bounds{Similar to 4-dimensional \ac{GUTs}~\cite{Agrawal:2022lsp}, the coupling to photons of any axion in the heterotic axiverse is constrained to be below the \ac{QCD} axion prediction~\cite{Agrawal:2024ejr,Reig:2025dqb}. }
\comments{Similar studies have been recently pushed forward for F-theory in Ref.~\cite{Fallon:2025lvn}.}

\subsubsection{High quality QCD axion from gauge symmetries \hfill\darkmatter\strongcp }
\purpose{Construct a \ac{QCD} axion model that naturally solves the strong CP problem without imposing an explicit global \ac{PQ} symmetry. In conventional \ac{PQ} models, the global \ac{PQ} symmetry can be explicitly broken by Planck-suppressed operators, thereby spoiling the “high quality” of the axion. In certain frameworks, however, the \ac{PQ} symmetry can emerge as an accidental symmetry of a gauge theory, thus being protected from symmetry-violating corrections.}
\modelling{A gauge structure of the form $G_{\text{SM}} \times G$ is introduced, where the effective \ac{PQ} symmetry arises from the gauge symmetry of $G$.  }
\paper{Possible realisations include $G$ being discrete gauge symmetries~\cite{Chun:1992bn,BasteroGil:1997vn}, Abelian gauge symmetries~\cite{Fukuda:2017ylt,Duerr:2017amf}, including a continuous supersymmetric R symmetry \cite{Unwin:2024yqq}, and non-Abelian gauge symmetries~\cite{Randall:1992ut,DiLuzio:2017tjx}.
}
\comments{
The gauge group typically needs to be sufficiently large in order to maintain the high quality of the \ac{PQ} symmetry for decay constants consistent with astrophysical bounds. 
If one introduces many additional gauge groups, one may have an effective theory of many axions, i.e., the axiverse~\cite{Lee:2018yak,Lee:2024xjb}.}

\subsubsection{Kaluza-Klein QCD Axions \hfill \stringtheory\strongcp }
\label{sec:KKaxions}
\purpose{Solve the strong CP problem dynamically by promoting the \ac{QCD} axion to an extra-dimensional bulk field .}
\modelling{The model features a \ac{QCD} axion which propagates in the bulk of extra dimensions, thus developing a \ac{KK}-tower in the resulting 4D \ac{EFT}. Different features can then be implemented in analogy to the other \ac{QCD} axion models.}
\paper{The ideas behind this model were studied in Refs.~\cite{Dienes:1999gw,deGiorgi:2024elx}.}
\useful{The general framework to study the impact of multiple axions in laboratory experiments can be found in Ref.~\cite{deGiorgi:2025ldc}.}
\bounds{Traditional axion bounds apply~\cite{AxionLimits}, but bounds need to be rescaled. An example of phenomenological application in the context of a flat extra dimension can be found in Ref.~\cite{Dienes:2012jb}.}
\comments{The same model can be implemented to generate approximate \hyperref[sec:maxions]{Maxions}.}
\subsubsection{KSVZ QCD Axion \hfill \darkmatter\strongcp }
\label{sec:KSVZ}
\purpose{Solve the strong CP problem dynamically.}
\modelling{The particle spectrum is enlarged by a complex singlet of the \ac{SM}, which is charged under the $U(1)_\text{PQ}$. This singlet couples to heavy vector-like fermions, which are charged under the gauge groups of the \ac{SM}. The interaction between the singlet and the \ac{SM} particles occurs through a loop mediated by the heavy vector-like fermions.}
\paper{The idea was first studied in Refs.~\cite{Kim:1979if,Shifman:1979if}.}
\useful{There are different valid charge assignments for the vector-like fermions, which lead to different phenomenology~\cite{DiLuzio:2024xnt}. A detailed discussion of the landscape of \ac{QCD} axion models beyond this one can be found in Ref.~\cite{DiLuzio:2020wdo}. Reference~\cite{GrillidiCortona:2015jxo} studies several properties of the \ac{QCD} axion at high precision, as well as the temperature dependency.}
\bounds{General \ac{ALP} limits apply, e.g.~\cite{AxionLimits}.}
\comments{Via the misalignment mechanism (see \ref{sec:misalignment}), it can be a viable \ac{DM} candidate~\cite{Abbott:1982af,Dine:1982ah,Preskill:1982cy}. It is subject to both the quality and \ac{DW} problems, although models with $N_\textrm{DW}=1$ can be easily found.}
\subsubsection{Majoron \hfill\darkmatter\neutrinos}
\label{sec:majoron}
\purpose{Goldstone boson associated with the spontaneous breaking of the global Lepton number, dynamically generating the Majorana mass scale for the seesaw mechanism. }
\modelling{Two typical realisations: (type-I) A scalar field coupled to the right-handed neutrino Majorana bilinear (type-II), the \ac{NGB} associated with the pseudoscalar component of the $SU(2)_L$ triplet. }
\paper{In the context of the type-I seesaw mechanism the model was firstly introduced in~\cite{Chikashige:1980qk,Chikashige:1980ui}, while for the type-II seesaw mechanism in~\cite{Gelmini:1980re}.}
\useful{An interesting Majoron model framework is the one presented in Ref.~\cite{Rothstein:1992rh}, where the global symmetry is protected against possible higher order terms in the potential.}
\bounds{Specific bounds come from \ac{DM}~\cite{Garcia-Cely:2017oco,Akita:2023qiz}, as well as Majoron couplings to right-handed neutrinos, which can lead to new signals at colliders~\cite{deGiorgi:2022oks,Marcos:2024yfm}}
\comments{The Majoron's weak couplings with \ac{SM} particles~\cite{Heeck:2019guh} make it a suitable \ac{DM} candidate, see the several works in this direction:~\cite{Berezinsky:1993fm,Reig:2019sok,Liang:2024vnd,Greljo:2025suh}. For type-II Majoron, see also Ref.~\cite{Biggio:2023gtm}. Some works addressing the mass of the Majoron can be found in Refs.~\cite{Frigerio:2011in,deGiorgi:2023tvn,Berbig:2025nrt}.
It is worth noting that low-scale Majorons are a popular candidate for alleviating the Hubble Tension~\cite{Escudero:2019gvw,Escudero:2021rfi,Fernandez-Martinez:2021ypo}.
}

\subsubsection{Maxions \hfill\strongcp}
\label{sec:maxions}
\purpose{Generate multiple axions displaced from the canonical \ac{QCD} axion band while solving the strong CP puzzle.}
\modelling{The model involves $N$ axions which couple to the $G\widetilde{G}$ term and, possibly, via an arbitrary potential. If the determinant of the $N$-axions mass matrix stemming solely from the external potential vanishes, the strong CP problem can be solved while simultaneously having the mass eigenmodes of the system displaced from the canonical \ac{QCD} band.}
\paper{The general framework and term were created in Ref.~\cite{Gavela:2023tzu}.}
\useful{A discussion about the impact of $N$ generic \ac{ALP}s in laboratory experiments such as light-shining-through-the-wall, helioscopes and haloscopes can be found in Ref.~\cite{deGiorgi:2025ldc}.}
\bounds{Traditional axion bounds~\cite{AxionLimits} apply, but need to be rescaled depending on the specific model.}
\comments{The features needed for the potential to generate maxions are non-generic. A spectrum which well approximates the maxions' features can be obtained, e.g.\ in extra-dimensional models (see \hyperref[sec:KKaxions]{KK QCD Axions}).}

\subsubsection{Minimal Axion Minimal Linear \texorpdfstring{$\sigma$}{sigma} Model \hfill\hierarchy\strongcp}
\purpose{Solve the Higgs hierarchy problem and the strong CP problem within the same construction.}
\modelling{The minimal $SO(5)/SO(4)$ linear $\sigma$ model, that represents a possible solution to the Higgs hierarchy problem, is extended including an additional
complex scalar field, singlet under the global \(SO(5)\) and the \ac{SM} gauge
symmetries. Two distinct possibilities for a \hyperref[sec:KSVZ]{KSVZ axion} arise: either it is a \ac{QCD} axion with an associated scale larger than $10^5$ TeV and therefore falling in the category of the invisible
axions; or it is a more massive \ac{ALP}, such as a 1 GeV axion with an associated scale of 1 TeV, that may show up in collider and flavour searches.}
\paper{Originally proposed in Ref.~\cite{Merlo:2017sun}.}
\useful{The Minimal Linear $\sigma$ Model has been first introduced and studied in Refs.~\cite{Feruglio:2016zvt,Aguilar-Saavedra:2019ghg}.}
\bounds{Bounds can be derived from astrophysics and cosmological sources for masses in the traditional \ac{QCD} axion region~\cite{Cadamuro:2011fd,Irastorza:2018dyq,AxionLimits}, or from colliders and flavour facilities for masses heavier than $0.1$ GeV~\cite{Bauer:2017ris,Bauer:2021mvw,MartinCamalich:2020dfe,Alda:2025uwo}.}
\comments{See Ref.~\cite{Brivio:2017sdm} for photonic signals in axion searches within this class of models. See also Ref.~\cite{Alonso-Gonzalez:2018vpc} for a specific realisation where the \ac{PQ} symmetry is \ac{SSB} at the TeV scale.}

\subsubsection{MFV Axion \hfill\flavour\strongcp}
\purpose{Applying the framework of \ac{MFV} to a \ac{QCD} axion/\ac{ALP}.}
\modelling{The \ac{SM} Lagrangian exhibits the \ac{MFV} symmetry, a $U(3)=SU(3)\times U(1)$ group for each fermion species, which decomposes into non-Abelian and Abelian factors. While the non-Abelian part governs the mass hierarchies within fermion families, the hierarchy between the top, bottom and tau can be explained using one of the Abelian factors. The core idea is to identify this Abelian factor with the \ac{PQ} symmetry, spontaneously broken by the \ac{VEV} of a complex scalar. The resulting axion/\ac{ALP} couples universally to fermions of the same species, but with distinct strengths for up-type quarks, down-type quarks and charged leptons.}
\paper{The  model is explained in~\cite{Arias-Aragon:2017eww}.}
\useful{Original papers on \ac{MFV}~\cite{DAmbrosio:2002vsn,Cirigliano:2005ck,Davidson:2006bd,Alonso:2011jd}.}
\bounds{The relevant phenomenology is determined by the mass range of the axion/\ac{ALP}. For small masses, the traditional astrophysical/cosmological bounds as well as terrestrial experiments on QCD-axions apply~\cite{Cadamuro:2011fd,Irastorza:2018dyq,AxionLimits}; otherwise, flavour and collider searches are relevant for larger masses~\cite{Bauer:2017ris,Bauer:2021mvw,MartinCamalich:2020dfe,Alda:2025uwo}.}
\comments{The original MFV Axion focuses on the quark sector, but the extension to the lepton sector could reduce the tension in the determination of the Hubble constant, $H_0$. This has been worked out in Ref.~\cite{Arias-Aragon:2020qip}.}

\subsubsection{Minimal Massive Majoron \hfill\neutrinos}
\purpose{Constructs a minimal framework where the smallness of neutrino masses and the mass of the Majoron share a common origin via the (low-scale) seesaw mechanism.}
\modelling{The model extends the \ac{SM} with sterile neutrinos and a complex scalar singlet. Unlike standard \ac{SSB} where the Majoron is a massless Goldstone boson, this framework incorporates a specific, explicit Lepton Number breaking term in the scalar potential. This same term is responsible for the suppression of light neutrino masses (Linear seesaw). As a result, the Majoron mass $m_J$ is not a free parameter but is strictly correlated with the light neutrino masses.}
\paper{A specific realisation was proposed in Ref.~\cite{deGiorgi:2023tvn}, building upon the symmetry structures analysed in Ref.~\cite{Frigerio:2011in}.}
\useful{For a broader review of the Linear seesaw mechanism and its distinction from the inverse seesaw, see Ref.~\cite{Lopez-Pavon:2012yda}.}
\bounds{The model can be tested by looking at the decay of the Majoron into \ac{SM} particles if it constitutes the total amount of \ac{DM} Ref.~\cite{Heeck:2019guh,Akita:2023qiz}. The astrophysical and cosmological bounds that apply to other light \ac{ALP}s also apply here~\cite{Cadamuro:2011fd,AxionLimits}.}

\subsubsection{Photophobic ALP\hfill\none}
\purpose{A scenario in which the \ac{ALP}--photon coupling is suppressed at tree level due to ultraviolet boundary conditions, leaving only loop-induced contributions. This allows a broad parameter region to evade conventional photon-based \ac{ALP} searches while remaining phenomenologically viable.}
\modelling{An \ac{ALP} with \ac{PQ} symmetry whose anomaly coefficients for the \ac{EW} gauge groups are arranged such that the $a\gamma\gamma$ coupling cancels at tree level, but it can have the coupling to the other gauge bosons. The coupling is dominantly with $W^+W^-$, $ZZ$ and $Z\gamma$.}
\paper{The theoretical framework and distinct phenomenology of this model were explored in Ref.~\cite{Craig:2018kne}}
\bounds{Constraints arise from \ac{EW} multi-boson final states at colliders, stellar cooling, supernova energy loss, and rare decay processes~\cite{Craig:2018kne}.}
\comments{Traditional photon-based \ac{ALP} searches, such as $a\!\to\!\gamma\gamma$ decays or photon-conversion experiments, lose sensitivity in this scenario. As a result, parameter regions that would be excluded in standard \ac{ALP} models remain open. Future collider measurements of triboson and diboson processes play an important role in probing this model.}

\subsubsection{Relaxion \hfill\hierarchy\strongcp}
\purpose{Dynamically solves the \ac{EW} hierarchy problem by promoting the Higgs mass squared to a field-dependent variable, allowing it to naturally relax from the cutoff scale to the observed weak scale.}
\modelling{An axion-like field $\phi$ rolls down a shallow slope $V_{roll} \sim g \Lambda^3 \phi$, scanning the Higgs mass parameter $\mu^2(\phi) \sim \Lambda^2 -g \phi$. When $\mu^2$ crosses zero, the Higgs acquires a \ac{VEV}, which turns on a periodic back-reaction potential $V_{br} \propto \Lambda_{br}^3 \langle h \rangle \cos(\phi/f)$. This barrier stops the rolling shortly after the \ac{EW} symmetry breaking. In the minimal model, this back-reaction is generated by \ac{QCD} instantons.}
\paper{Originally proposed in Ref.~\cite{Graham:2015cka}.}
\useful{The minimal model typically yields $\theta_{\text{QCD}} \sim 1$ (violating CP). Extensions ensuring $\theta_{\text{QCD}} \approx 0$ to solve the strong CP problem simultaneously are discussed in Ref.~\cite{Nelson:2017cfv,Chatrchyan:2022pcb}.}
\bounds{Unlike a standard generic \ac{ALP}, the Relaxion mixes with the Higgs boson. Consequently, it inherits scalar couplings to matter. Bounds arise from fifth-force searches, stellar cooling and rare meson decays, as analysed in Refs.~\cite{Flacke:2016szy,Frugiuele:2018coc,Chatrchyan:2022dpy}. Standard axion limits apply~\cite{AxionLimits}.}
\comments{A major theoretical challenge is the requirement for an enormous number of inflationary e-folds to allow the field to scan the full parameter space. This imposes constraints on the inflation sector (e.g.\ requiring low-scale inflation). These constraints can be weakened when including other sources of friction beyond Hubble friction, for instance through relaxion fragmentation~\cite{Fonseca:2019lmc}. In addition, an alternative stopping mechanism exists when considering quantum fluctuations of the relaxion \cite{Chatrchyan:2022pcb}, which also opens the possibility for the relaxion to be dark matter~\cite{Chatrchyan:2022dpy}.} 
\subsubsection{Relaxion for cosmological constant \hfill\hierarchy}
\purpose{The model was created to address the smallness of the present cosmological constant.}
\modelling{The cosmological constant is dynamically scanned to small values by the slow roll of a scalar field (the relaxion) with an extremely flat potential, assuming that a vanishing cosmological constant is critical for cosmological evolution.}
\paper{Originally proposed in Refs.~\cite{Abbott:1984qf,Banks:1984tw}.}
\useful{Modern realisation with bouncing cosmology~\cite{Graham:2019bfu,Ji:2021mvg} or specific axion-like inflaton potentials~\cite{Yin:2021uus}.}
\bounds{Constraints depend on the relaxion mass and couplings~\cite{AxionLimits}, and specifically on the equation of state of the present Universe~\cite{Graham:2019bfu,Ji:2021mvg,Yin:2021uus}.}
\comments{Some scenarios predict time-varying \ac{DE}, potentially consistent with recent DESI data~\cite{Yin:2024hba}. However, achieving arbitrarily efficient relaxation without entering a regime of eternal inflation remains a significant model-building challenge.}


\subsubsection{SM*A*S*H \hfill \darkmatter\inflation\neutrinos\strongcp}
\label{sec:SMASH}
\purpose{SM*A*S*H (``Standard Model*Axion*Seesaw*Higgs-Portal Inflation'') was built to simultaneously address vacuum stability, inflation, baryon asymmetry, neutrino masses, strong CP problem and \ac{DM}.}
\modelling{The field content is comprised of the \ac{SM} augmented with a complex \ac{PQ}-charged singlet, vector-like fermions and three (gauge singlet) right-handed neutrinos, with specific \ac{PQ} charges.}
\paper{The model was proposed and named in Refs.~\cite{Ballesteros:2016euj,Ballesteros:2016xej}.}
\useful{A review of the model, as well as some extensions, was presented in Ref.~\cite{Ballesteros:2019tvf}. An overview of experimental prospects to further constrain SM*A*S*H (\ac{CMB} polarisation, axion haloscopes, space-borne \ac{GW} detectors) can be found in Ref.~\cite{Ringwald:2023anj}.}
\bounds{Constraints can be found in Ref.~\cite{Ballesteros:2016euj,Ballesteros:2016xej,Ballesteros:2019tvf}.}
\comments{Similar models have been considered in Refs.~\cite{Sopov:2022bog,Ballesteros:2019tvf,Berbig:2022pye,Matlis:2023eli,Boucenna:2014uma}. The \ac{PQ} scale links axion \ac{DM} and heavy-neutrino masses, enabling seesaw neutrino masses and thermal leptogenesis.}

\subsubsection{Supersymmetric Nonlinear Sigma Model Axion \hfill\darkmatter\strongcp}
\purpose{Simultaneously addresses the gauge hierarchy problem and the origin of quark/lepton generations by identifying sfermions as \ac{NGB} on a specific coset space.}
\modelling{The Higgs and matter supermultiplets are embedded into a nonlinear sigma model on a Kähler manifold (such as the coset $E_7/SU(5)$). In this framework, the $U(1)$ symmetry associated with the manifold's Kähler potential is typically anomalous. To restore consistency in supergravity, a specific multiplet containing an axion must be introduced to cancel this anomaly. This axion naturally acquires a coupling to QCD, functioning as the \ac{QCD} axion. }
\paper{The consistency conditions requiring the axion in these \ac{SUGRA} frameworks were derived in Refs.~\cite{Komargodski:2010rb,Kugo:2010fs}. Ref.~\cite{Yanagida:2019evh} specifically identifies this required field as the \ac{QCD} axion and studies its viability as \ac{DM}.}
\bounds{Standard \ac{ALP} limits apply~\cite{AxionLimits}. Additionally, bounds on the supersymmetric spectrum are non-standard, as the sfermions (being \ac{pNGB}s) may have compressed spectra or specific mass relations distinct from \ac{MSSM} benchmarks.}
\comments{A compelling feature is the potential explanation for the three generations of matter. This arises from the topology of the coset space $G/H$, where the number of families is determined by topological invariants of the manifold~\cite{Kugo:1983ai,Yanagida:1985jc}. 
}

\subsubsection{Type IIB String Theory Axiverse \hfill\stringtheory}
\purpose{Investigates the ``axiverse", the scenario with many light \ac{ALP}s within the framework of Type IIB string theory.}
\modelling{\ac{ALP}s emerge naturally from the complex topology of the extra dimensions (Calabi-Yau manifolds). By analysing large datasets of these geometric shapes, one can statistically predict the mass spectrum and couplings of the resulting population of axions.}
\paper{A statistical survey of these models is presented in Ref.~\cite{Gendler:2023kjt}.}
\useful{The term ``axiverse'' was originally coined in Ref.~\cite{Arvanitaki:2009fg}.}
\bounds{Constraints apply to the axion-photon couplings~\cite{Gendler:2023kjt} and the mass spectrum via black hole superradiance~\cite{Mehta:2021pwf}. Cosmological limits from the \ac{CMB} and Lyman-$\alpha$ forest also constrain the underlying geometry~\cite{Jain:2025vfh}.}


\subsubsection{Variant Axion Models \hfill\darkmatter\strongcp }
\purpose{To construct axion models without the domain wall problem beyond the minimal \hyperref[sec:DFSZ]{DFSZ} frameworks by allowing generation- or flavour-dependent \ac{PQ} charge assignments, while preserving the solution to the strong CP problem.}
\modelling{The axion couples non-universally to quarks and leptons through flavour-dependent \ac{PQ} charges. Such “variant” models modify the structure of axion couplings compared to standard models, leading to different phenomenology in hadronic, leptonic and flavour sectors.}
\paper{The model was proposed in Ref.~\cite{Bardeen:1986yb}.}
\bounds{Constrained by standard astrophysical axion limits~\cite{AxionLimits}, as well as by rare meson decays and flavor-violating processes~\cite{Bardeen:1986yb,MartinCamalich:2020dfe}. 
}
\comments{Provides a general framework for axion models with flavour-dependent couplings. The models are still relevant today in the context of axion-flavour physics and motivate searches for flavour-violating decays involving axions.}

\subsubsection{\texorpdfstring{$\mathcal{Z}_N$}{ZN} Axions \hfill\strongcp}
\purpose{Solves the strong CP problem dynamically while allowing the axion to be exponentially lighter than the canonical \ac{QCD} axion prediction ($m_a \ll m_{\pi} f_{\pi}/f_a$).}
\modelling{The model features $N$ copies of the \ac{SM} linked by the
axion field satisfying a discrete $\mathcal{Z}_N$ symmetry. For the axion, the symmetry is implemented non-linearly. 
The \ac{PQ} symmetry is broken by the anomalies of all the \ac{QCD} gauge group copies, generating different contributions to the axion potential. However, these contributions exhibit an exponential cancellation due to the $\mathcal{Z}_N$ symmetry.}
\paper{Originally proposed in Ref.~\cite{Hook:2018jle} and further developed in Ref.~\cite{DiLuzio:2021pxd}.}
\bounds{A set of constraints coming from a variety of experiments can be found in Ref.~\cite{DiLuzio:2021pxd}.}
\comments{As explored in the original idea~\cite{Hook:2018jle}, the same mechanism that suppresses the axion potential can be used to solve the (little) hierarchy problem.}
\subsection{Scalar Models}
\subsubsection{\texorpdfstring{$\alpha$}{alpha}-Attractors \hfill\inflation}
\label{Sec:alpha_attractor}
\purpose{To provide a broad class of inflationary models with universal predictions for the spectral index ($n_s$) and tensor-to-scalar ratio ($r$), largely independent of the specific potential, while naturally featuring an inflaton that behaves as a \ac{WISP}.}
\modelling{The inflaton is a scalar field with a kinetic term that has a pole in the field space, leading to an ``attractor" behaviour in the cosmological observables. The Lagrangian is characterised by a parameter $\alpha$ related to the curvature of the Kähler manifold in supergravity embeddings. In the Einstein frame, this results in a stretched, flat potential that makes the inflaton behave as a \ac{WISP} during the inflationary epoch, with specific, predictive self-interactions.}
\paper{The framework of cosmological attractors and their embedding in supergravity were established in Refs.~\cite{Kallosh:2013hoa,Ferrara:2013rsa,Kallosh:2013yoa}.}
\useful{Comprehensive review:~\cite{Alho:2017opd}. A connection to conformal symmetry and supergravity can be found in ref.~\cite{Kallosh:2013hoa}. Constraints on reheating and freeze-in \ac{DM}~\cite{Ghoshal:2025ejg}.}
\bounds{Predictions are in excellent agreement with \emph{Planck} data~\cite{Planck:2018vyg} and are prime targets for future \ac{CMB} experiments (e.g., Simons Observatory, LiteBIRD)~\cite{SimonsObservatory:2018koc, Hazumi:2019lys} .
The tensor-to-scalar ratio is bounded by $r \approx 3\alpha/N^2$ for $N$ e-folds, typically giving $10^{-3} \lesssim r \lesssim 10^{-2}$ for $\alpha \sim \mathcal{O}(1-10)$.}
\comments{The model provides a compelling connection between supergravity/string theory constructions and observable inflation. The parameter $\alpha$ is related to fundamental theory: $\alpha=1$ corresponds to conformal models, $\alpha=2$ to fiber inflation (cf. Sec.~\ref{Sec:fiber}), while $\alpha\to\infty$ recovers the $\mathcal{R}^2$ (Starobinsky) limit, cf.~\ref{subsec:R2}. The attractor mechanism makes the inflationary predictions robust against quantum corrections.}

\subsubsection{Axio-dilaton \hfill\darkenergy\darkmatter\stringtheory}
\purpose{The axio-dilaton is one of the fields arising in string theory and can also realise \ac{DM} and quintessence \ac{DE}.}
\modelling{The axio-dilaton is a complex field combining the axion and \hyperref[sec:dilaton]{dilaton} fields. In string theory, it is the scalar part of the supermultiplet that contains the dilaton. The joint dynamics of the axion and dilaton allow the axio-dilaton to evade solar system fifth force constraints and realise quintessence \ac{DE}~\cite{Burgess:2021obw}. The axio-dilaton system can realise both \ac{DM} and \ac{DE}~\cite{Smith:2024ibv}.}
\paper{The foundational work can be found in Ref.~\cite{Shapere:1991ta}.}
\useful{A useful review paper covering the axion-dilaton and other topics in string cosmology is Ref.~\cite{Cicoli:2023opf}.}
\subsubsection{Chameleons \hfill\darkenergy \darkmatter \inflation }
\label{sec:Chameleons}
\purpose{Scalar fields that evolve cosmologically so that they can have $\mathcal{O}(1)$ couplings to matter, instead of unnaturally small couplings to satisfy constraints from the equivalence principle.}
\modelling{Chameleons have non-linear self-interactions and a conformal or disformal coupling to the metric tensor, resulting in an effective mass depending on the local matter density. In denser environments, chameleons are more massive, which reduces the range of chameleon fifth forces and is the essence of the chameleon screening mechanism. }
\paper{Refs.~\cite{Khoury:2003rn,Khoury:2003aq,Brax:2004qh} introduced chameleons and discussed their role as a quintessence model.}
\useful{A detailed introduction, including the effective potentials, can be found in Ref.~\cite{Burrage:2016bwy}. Chameleons have been discussed in the context of inflation, see for instance Refs.~\cite{Hinterbichler:2013we,Saba:2017xur,Bernardo:2017xcm}, and as \ac{DM} candidates in Refs.~\cite{Katsuragawa:2016yir,Katsuragawa:2017wge,Chen:2019kcu,Zaregonbadi:2025ils}.}
\bounds{All currently existing bounds on chameleons can be found in Refs.~\cite{Burrage:2017qrf,Fischer:2024eic} and in references therein.}
\comments{Modelling must consider bounds coming from the equivalence principle as well as from fifth force experiments. Note that, for some parameter values, chameleon modified gravity is equivalent to $f(R)$-gravity~\cite{Nojiri:2017ncd}, scale-dependent gravity~\cite{Neckam:2025kip} or Higgs portal models~\cite{Burrage:2018dvt}.}

\subsubsection{Curvaton \hfill\inflation}
\purpose{To generate the observed primordial density perturbations from a spectator field rather than the inflaton, allowing for inflation models that would otherwise be ruled out by spectral tilt or normalisation constraints.}

\modelling{The curvaton is a prototypical \ac{WISP} candidate: a light scalar field ($\sigma$) that is subdominant during inflation but acquires quantum fluctuations. To generate a scale-invariant spectrum, the curvaton must be effectively massless during inflation ($m_\sigma \ll H$), satisfying the ``Slim" criteria. After inflation, it oscillates, dominates the energy density (or decays) and converts its entropy perturbations into the adiabatic curvature perturbations ($\zeta$) observed in the \ac{CMB}.}

\paper{Some foundational works about the model can be found in Refs.~\cite{Lyth:2001nq,Moroi:2001ct,Enqvist:2002rf}.}

\useful{A review of non-Gaussianity generation can be found in Ref.~\cite{Bartolo:2004if}. Ref.~\cite{Dimopoulos:2003az} presents the curvaton as a \ac{pNGB}. Ref.~\cite{Takahashi:2021bti} analyses what are the cosmological parameters suggested by solutions to the Hubble tension based on an \ac{ALP} curvaton.
}

\bounds{The model is strongly constrained by \emph{Planck} limits~\cite{Planck:2018vyg} on local non-Gaussianity ($f_{\text{NL}}$) and isocurvature perturbations. It requires the curvaton to decay before \ac{BBN} to avoid spoiling the correct light element abundances.}

\comments{Curvatons can revive ``dead" inflation models (like low-scale inflation). The curvaton is naturally identified with an \ac{ALP} or a \ac{pNGB}; this identification provides a shift symmetry that protects the required light mass from radiative corrections, creating a direct link between the \ac{WISP} landscape and the generation of cosmic structure. A key signature is large non-Gaussianity ($f_{\text{NL}} \sim 1/r_D$, where $r_D$ is the decay fraction), which is much larger than the standard single-field prediction.}

\subsubsection{D-BIon \hfill\darkenergy}
\purpose{The model was introduced as a possible \ac{DE} candidate that has a screening mechanism in order to avoid Solar System constraints on scalar fifth forces.}
\modelling{The model is based on the Dirac-Born-Infeld action, but uses a flipped sign both for the total action and in front of the kinetic term in order to achieve screening.}
\paper{D-BIonic screening was first briefly discussed in Ref.~\cite{Dvali:2010jz} and later fully developed in Ref.~\cite{Burrage:2014uwa}.}
\useful{Short and useful introductions to the model as can be found in Refs.~\cite{Panpanich:2017nft,BeltranJimenez:2024zmd}.} 
\bounds{Solar System constraints on this model are discussed in Ref.~\cite{Burrage:2014uwa}.}

\subsubsection{D-Brane Inflation  \hfill \inflation\stringtheory}
\label{subsubsec:D-Brane}
\purpose{To realise inflation within string theory by treating the position of a D-brane in the compactified extra dimensions as the inflaton field, which naturally behaves as a \ac{WISP} due to the shift symmetry of its position modulus.}
\modelling{The inflaton is identified with the scalar field parameterising the radial position of a D3-brane moving in a warped throat region of a Type IIB string compactification. The potential arises from the brane's tension and its interaction with background fluxes and other branes. The flatness of the potential is protected by the approximate shift symmetry of the brane position, making it a natural \ac{WISP} candidate. In the \ac{UV}-complete string theory context, the model predicts specific relationships between the inflationary observables and the properties of the compactification.}
\paper{The detailed construction of the inflaton in warped geometries is detailed in Ref.~\cite{Kachru:2003aw}, while specific contributions from non-perturbative effects and brane-antibrane forces were rigorously determined by Ref.~\cite{Baumann:2006th} and Ref.~\cite{Bena:2010ze}, respectively.}

\useful{Comprehensive reviews on string inflation and the D-brane construction:~\cite{Baumann:2009ds, Baumann:2014nda}. Connection to the axiverse and multiple light fields:~\cite{Arvanitaki:2009fg}.}
\bounds{\ac{CMB} constraints on the scalar spectral index  (\(n_s\)) and tensor-to-scalar ratio (\(r\)) by \textit{Planck}~\cite{Planck:2018vyg}, \emph{Planck}+ACT~\cite{ACT:2025tim}, and future observations~\cite{SimonsObservatory:2018koc, Hazumi:2019lys}. The warped scale $\Lambda_{\mathrm{UV}}$ is constrained by the amplitude of density perturbations. Cosmic string constraints from the D-brane annihilation phase:~\cite{Sarangi:2002yt, Copeland:2003bj}.}
\comments{The model naturally connects inflation to other string-theoretic \ac{WISP}s. After inflation, the D-brane position modulus can become an ultralight \ac{ALP}, contributing to the axiverse. The end of inflation typically involves brane-antibrane annihilation, producing cosmic strings that are constrained by \ac{CMB}~\cite{Planck:2018vyg, SimonsObservatory:2018koc, Hazumi:2019lys, BICEP:2021xfz, ACT:2025tim} and gravitational wave observations~\cite{Auclair:2019wcv}. The reheating temperature is set by the decay into closed string modes (gravitons, moduli) and \ac{SM} particles living on other branes.}

\subsubsection{Dilaton \hfill \darkmatter\stringtheory}
\label{sec:dilaton}
\purpose{The dilaton $\phi$ is a fundamental scalar field appearing in the massless spectrum of all string theories. Its vacuum expectation value dynamically determines the string coupling constant ($g_s \propto \exp{\phi}$). It is the scalar partner to the graviton and often plays a role in stabilising the volume of compact dimensions.}
\modelling{A massless dilaton mediates a ``fifth force" and induces variations in constants. To evade detection, the field dynamics must minimise these couplings; there exist mechanisms which provide a cosmological attractor that drives the field towards a regime where it decouples from the \ac{SM}~\cite{Damour:1994zq}. If the dilaton acquires a mass (e.g., via supersymmetry breaking), it can serve as a cold \ac{DM} candidate produced via the misalignment mechanism (see \ref{sec:misalignment}).}
\paper{The string origin is detailed in Ref.~\cite{Witten:1984dg}, while the screening mechanism is established in Ref.~\cite{Damour:1994zq}.}
\bounds{Stringent bounds arise from tests of the Equivalence Principle with (MICROSCOPE~\cite{Touboul:2017grn}), Solar System gravity tests (Cassini~\cite{Bertotti:2003rm}), and constraints on the time-variation of the fine-structure constant~\cite{Damour:2010rm}.} 
\subsubsection{Environment-dependent dilaton \hfill\darkenergy\stringtheory }
\purpose{String theory-inspired scalar field model that was proposed as a possible candidate for \ac{DE} and which circumvents Solar System-based fifth force constraints due to a screening mechanism.}
\modelling{Its potential naturally arises in the strong coupling limit of string theory (see also \hyperref[sec:dilaton]{Dilaton}). Depending on the considered parameter regime, the dilaton fifth force is screened in sufficiently dense environments either due to the Damour-Polyakov mechanism or the chameleon mechanism~\cite{Fischer:2023koa} (see \hyperref[sec:Chameleons]{Chameleons}).}
\paper{The Damour-Polyakov mechanism was proposed in Ref.~\cite{Damour:1994zq} and the environment-dependent dilaton model with cosmological applications was developed in Refs.~\cite{Gasperini:2001pc,Damour:2002nv,Damour:2002mi,Brax:2010gi,Brax:2011ja}.}
\useful{A theoretical discussion of the model in laboratory and astrophysical environments, including analytical solutions for the 1D equation of motion in the presence of one mirror, two mirrors or  a sphere, can be found in Ref.~\cite{Brax:2022uyh}.}
\bounds{All currently existing experimental bounds from tests in laboratories and in the Solar System can be found in Ref.~\cite{Fischer:2024eic}.}

\subsubsection{Fibre Inflation \hfill \stringtheory \inflation}
\label{Sec:fiber}
\purpose{To realise large-field inflation within Type IIB string theory compactifications using Kähler moduli as the inflaton, providing a \ac{UV}-complete model with specific predictions and connecting to the broader axiverse paradigm.}
\modelling{The inflaton is a specific combination of Kähler moduli (specifically a ``fibre modulus") in a Calabi-Yau compactification with a large-volume scenario. The potential arises from perturbative corrections to the Kähler potential ($\alpha'$ and string loop effects) that break the no-scale structure and generate a plateau-like potential. The inflaton is a \ac{WISP} with gravitational-strength couplings, protected by the approximate shift symmetry of the modulus field.}
\paper{The model was originally proposed in Ref.~\cite{Cicoli:2008va}.}
\useful{A comprehensive review of string inflation moduli stabilisation can be found in Ref.~\cite{Cicoli:2023opf}, while a connection to the axiverse and dark radiation in Ref.~\cite{Cicoli:2010ha}. Detailed phenomenological analysis and \ac{CMB} predictions:~\cite{Cicoli:2008gp}.}
\bounds{Predicts a scalar spectral index (\(n_s\))  and tensor-to-scalar ratio (\(r\)) consistent with \emph{Planck}/BICEP/\emph{Keck} data~\cite{Planck:2018vyg, BICEP:2021xfz}. The model requires a high string scale $M_s \sim 10^{15}$ GeV and generates observable primordial gravitational waves. The decay to hidden sector axions can produce dark radiation with $\Delta N_{\text{eff}} \sim 0.5$-$1$, testable with \ac{CMB} future probes ~\cite{SimonsObservatory:2018koc, Hazumi:2019lys}.}
\comments{Represents a concrete, \ac{UV}-complete realisation of inflation in string theory where the inflaton is unequivocally a \ac{WISP} (a light modulus). The model naturally connects to the axiverse through the presence of multiple light axions from the same compactification. Reheating occurs through gravitational particle production and modulus decays to both visible and hidden sectors, potentially leaving imprints in dark radiation. The framework is particularly robust as it does not require tuning of the superpotential. Phenomenologically, the model falls into the class of cosmological \(\alpha\)-attractors with characteristic parameter \(\alpha=2\), making robust predictions for the spectral index (\(n_s\)), and the tensor-to-scalar ratio (\(r\)) largely independent of the potential's fine details, cf. Sec.~\ref{Sec:alpha_attractor}.}
\subsubsection{Galileons \hfill\darkenergy \inflation}
\label{sec:Galileons}
\purpose{Proposed as possible explanations for the accelerated expansion of the Universe, while avoiding Solar System constraints on scalar fifth forces due to a screening mechanism.}
\modelling{Galileon models arise, e.g., from the \ac{DGP} model~\cite{Dvali:2000rv}. It is assumed that the theory obeys a covariant generalisation of the Galileon shift symmetry~\cite{Burrage:2010zz}. Having some of the non-canonical kinetic terms known from Horndeski theory~\cite{Horndeski:1974wa}, Galileon models screen their fifth forces in dense environments through the Vainshtein mechanism~\cite{Vainshtein:1972sx}.}
\paper{The Galileon shift symmetry first appeared in the context of the \ac{DGP} model~\cite{Dvali:2000hr}. Galileons as a scalar field model were then introduced in Ref.~\cite{Nicolis:2008in}.}
\useful{A review of Galileons in the contexts of gravitation and cosmology can be found in~\cite{deRham:2012az}. Discussions in the context of inflation can be found in Refs.~\cite{Kobayashi:2010cm,Burrage:2010cu}. For a specific inflationary application, see Sec.~\ref{sec:Non_min_Gal}.} 
\bounds{Most constraints on these models stem from astrophysics or cosmology; see, for example, Refs.~\cite{DeFelice:2011aa,Ali:2012cv,Jamil:2013yc,Neveu:2014vua,Neveu:2016gxp,Bartlett:2020tjd,Lawrence:2020mmu,Ferrari:2023qnh}. Due to the Vainshtein mechanism, it is much more intricate to constrain Galileons with Earth-based experiments, but there have been discussions about laboratory tests in Ref.~\cite{Brax:2011sv}. There also exist theoretical constraints from positivity bounds~\cite{Melville:2022ykg}.}

\subsubsection{Generalised Symmetrons \hfill\darkmatter}
\label{sec:GenSymmetrons}
\purpose{Generalisations of the archetypal symmetron model to higher power potentials (see \hyperref[sec:Symmetron]{Symmetron}).}
\modelling{All generalised symmetron models behave qualitatively similar to the original symmetron, i.e., their fifth forces are screened by the Damour-Polyakov mechanism realised through the restoration of the potentials' $\mathbb{Z}_2$ symmetry in dense environments.}
\paper{Models were discovered using tomographic
methods in Refs.~\cite{Brax:2011aw,Brax:2012gr}.}
\useful{A short introduction to these models and a discussion of their fifth forces as an alternative explanation for the observed difference between baryonic and lens masses of galaxies can be found in Ref.~\cite{Kading:2023hdb}.}
\bounds{Currently, there exist no published bounds on any generalised symmetron model beyond the archetypal case.}
\comments{These models have barely been studied. An interesting difference between the archetypal symmetron and its generalisations is that, in environments with densities sufficiently high for total fifth force screening, nearly all of the latter have vanishing masses (see Ref.~\cite{Kading:2023hdb}).}

\subsubsection{Geometric Moduli \hfill\stringtheory}
\purpose{Geometic moduli arise in string theory models, corresponding to variation in the size of the compactified dimensions.}
\modelling{Extensive theoretical work has been undertaken to stabilise string moduli so they do not conflict with cosmology~\cite{McAllister:2023vgy}. Consistent string phenomenology requires that the moduli have potentials, as massless moduli generate fifth forces excluded by existing observations. Foundational work on moduli stabilisation can be found in Ref.~\cite{Kachru:2003aw}. In Ref.~\cite{Balasubramanian:2005zx}, it is demonstrated that moduli stabilisation can be achieved in the limit of large volumes of the compactification manifold.}
\paper{The foundational works can be found in Refs.~\cite{Binetruy:1986ss,Witten:1984dg}. The first use of the term `moduli' known to the authors in this context is in~\cite{Damour:1995zq}.}
\bounds{Variations in the geometric moduli lead to variations in fundamental constants of the \ac{SM}, with bounds arising from searches for these variations~\cite{Damour:2010rm}.}

\subsubsection{Higgs Inflation \hfill\hierarchy \inflation}
\purpose{To identify the \ac{SM} Higgs boson as the inflaton, providing a minimal connection between collider physics and early-Universe cosmology.}
\modelling{In this model, the \ac{SM} Higgs boson plays the role of the inflaton, the field responsible for driving cosmic inflation in the early Universe. To achieve this, the Higgs field is allowed to interact with gravity in a ``non-minimal" way, meaning there's an extra term in the theory that links the Higgs directly to the curvature of spacetime (represented by the Ricci scalar $\mathcal{R}$). This coupling is parameterised by a constant $\xi$, so the interaction term is \(\xi |H|^2\mathcal{R}\). The exact details of how inflation unfolds depend on the underlying formulation of gravity, whether we use the standard ``metric" approach (where the metric is the fundamental variable) or the ``Palatini" approach (where the connection is treated independently), leading to different predictions and constraints.}

\paper{The proposal to identify the \ac{SM} Higgs boson as the inflaton was first presented in Ref.~\cite{Bezrukov:2007ep}.}
\useful{Review of Higgs inflation variants~\cite{Rubio:2018ogq}. Unitarity analysis~\cite{Bezrukov:2010jz, Ito:2021ssc}. Criticality of the \ac{SM}~\cite{Hamada:2014iga}.}
\bounds{In the \emph{Metric Formalism:} the model requires $\xi \sim 10^4$, it predicts tensor-to-scalar ratio $r \approx 0.003$ and spectral index $n_s \approx 0.965$, consistent with current data but facing unitarity issues at scale $\Lambda \sim M_{\rm Pl}/\xi$. In the \emph{Palatini Formalism:} the model requires $\xi \sim 10^9$. Predicts a heavily suppressed tensor-to-scalar ratio $r \sim 10^{-12}$ and slightly different $n_s$, with a higher unitarity cutoff $\Lambda \sim M_{\rm Pl}/\sqrt{\xi}$~\cite{Barbon:2009ya}. However, unlike the metric case, these predictions are highly sensitive to corrections to the potential, such as higher-dimensional operators or $\mathcal{R}^2$ terms.}

\comments{
The Mixed Higgs-$\mathcal{R}^2$ Inflation adds an $\mathcal{R}^2$ term to the action. In the metric formulation, this introduces a scalaron that ``unitarises" the theory up to the Planck scale~\cite{Granda:2019wip,Kim:2025ikw}. In the Palatini formulation, this framework has been embedded into \ac{GUTs}, highlighting a complementarity between \ac{CMB} observables and proton decay searches~\cite{Bostan:2025jkt}.
Analysis is typically performed in the Einstein framework after a conformal transformation to flatten the potential. In the unitary gauge, the Higgs phase is fixed, leaving a single real scalar inflaton.  Furthermore, vacuum stability requires the Higgs self-coupling to remain positive up to the inflationary scale, making the model critically dependent on the precise values of the top quark mass and strong coupling constant.}

\subsubsection{Kähler Moduli Inflation \hfill\hierarchy\inflation\stringtheory}
\purpose{To create a model of cosmic inflation using ideas from Type IIB string theory, where the inflaton is a special kind of scalar field called a Kähler modulus. This modulus controls the ``size" or volume of the extra, hidden dimensions in the theory, and it naturally acts like a \ac{WISP} because its interactions are very weak (only at the strength of gravity) and protected by a symmetry that keeps its potential flat and stable. }
\modelling{In this setup, the inflaton is the Kähler modulus, which is essentially a scalar field that describes the size of certain 4-dimensional cycles (like loops or surfaces) within a complex geometric space called a Calabi-Yau manifold, used in string theory to compactify (or ``curl up") extra dimensions. The potential energy for this field comes from subtle, non-perturbative effects, such as those from special branes (Euclidean D3-branes) or the condensation of gauginos (supersymmetric partners of gauge bosons) on D7-branes. These effects, combined with small corrections from quantum gravity (known as $\alpha'$ corrections), create a flat ``race-track" shaped potential, like a gently sloping track that allows the field to roll slowly, sustaining inflation. The Kähler modulus qualifies as a classic \ac{WISP} because its couplings to \ac{SM} particles (which live on D-branes in the larger space) are suppressed by the Planck scale, meaning they're extremely feeble.}
\paper{Early proposals linking Kähler moduli to quintessence and inflation include Refs.~\cite{Hellerman:2001yi, Brustein:2002mp}, while the specific Kähler Moduli Inflation scenario was formalised in Ref.~\cite{Conlon:2005jm}.}
\useful{Comprehensive reviews on string inflation and moduli stabilization:~\cite{Baumann:2014nda, Cicoli:2023opf, Chernoff:2014cba, Quevedo:2002xw}. Analysis of non-Gaussianities and preheating in these scenarios:~\cite{Adshead:2011jq, Wang:2011ed}. Connection to the broader axiverse:~\cite{Arvanitaki:2009fg}. Related brane inflation frameworks:~\cite{Dvali:2001fw, Shiu:2001sy, Choudhury:2003vr}.}
\bounds{\ac{CMB} constraints require the scalar spectral index (\(n_s\)) and the tensor-to-scalar ratio (\(r\)) to be within a certain range, which is consistent with \emph{Planck} data~\cite{Planck:2018vyg} and other experiments~\cite{SimonsObservatory:2018koc, Hazumi:2019lys, BICEP:2021xfz, ACT:2025tim}. The moduli masses are constrained by cosmological moduli problems to be $m_T \gtrsim 30$ TeV to avoid disrupting \ac{BBN}~\cite{Kawasaki:1994sc,Kawasaki:2004qu, Lemoine:2009is}. The compactification volume $\mathcal{V}$ is typically large $\mathcal{V} \sim 10^6-10^7$ in string units, suppressing the inflationary scale~\cite{Conlon:2005jm, Baumann:2014nda, Cicoli:2008va}.}
\comments{It represents a top-down approach to inflation where the \ac{WISP} inflaton emerges naturally from string theory. The model requires careful moduli stabilisation, often achieved via the Large Volume Scenario. After inflation, the Kähler modulus can decay gravitationally, reheating the Universe to temperatures $T_{\text{reh}} \sim 1-10$ GeV, potentially enabling non-thermal \ac{DM} production. The specific non-Gaussianity signature $f_{\text{NL}} \sim \mathcal{O}(1-10)$ provides a key testable prediction distinguishing it from single-field models. For a related framework in string-theoretic inflation, see \ref{subsubsec:D-Brane}.} 

\subsubsection{Non-Minimal Covariant Galileon Inflation \hfill\inflation}
\label{sec:Non_min_Gal}
\purpose{To build a specific inflationary model using the Galileon framework (see Sec.~\ref{sec:Galileons}), incorporating special self-interactions for the inflaton field protected by Galileon symmetry, while ensuring alignment with gravitational wave observations.}
\modelling{This model applies the Galileon framework to inflation, using a scalar field with higher-order derivative interactions that remain invariant under the Galileon shift symmetry. The inflaton features non-minimal coupling to gravity, which prevents ghost instabilities and enables a self-accelerated de Sitter-like expansion phase. The Galileon symmetry naturally protects the inflaton's mass from large quantum corrections, making it a technically natural \ac{WISP} candidate.}
\paper{The Galileon framework was introduced in Ref.~\cite{Nicolis:2008in}, adapted to inflation in Ref.~\cite{Creminelli:2010qf}, and extended to potential-driven scenarios in Ref.~\cite{Ohashi:2012wf}.}
\useful{For the general Galileon framework, see \hyperref[sec:Galileons]{Galileons}. A connection to Horndeski theories is discussed in~\cite{Kobayashi:2011nu}, a review of cosmological implications is~\cite{Deffayet:2013lga} and constraints from gravitational waves in~\cite{Lombriser:2016yzn}.}
\bounds{Severely constrained by the near-luminal propagation of gravitational waves (GW170817)~\cite{Baker:2017hug}, which rules out most pure covariant Galileon models. Viable parameter space requires significant tuning or extension beyond the minimal setup.}
\comments{This inflationary application of the Galileon framework demonstrates how derivative self-interactions can be used to construct inflation models, though the minimal version is largely ruled out by gravitational wave observations. It provides the foundation for more sophisticated \ac{WISP} constructions in Horndeski and beyond-Horndeski theories.}
\subsubsection{Natural Inflation \hfill\darkmatter\inflation}
\purpose{To provide a well-motivated inflationary model where the flatness of the potential is protected by a shift symmetry, as expected for an axion-like particle.}
\modelling{The inflaton is a \ac{pNGB} (axion) with a periodic potential. The symmetry protection makes the model technically natural, and the predictions are determined by the decay constant $f$. For $f \gtrsim M_{\mathrm{Pl}}$, the model produces \ac{CMB}-compatible predictions.}
\paper{The model was originally proposed in Ref.~\cite{Freese:1990rb} and further developed in Ref.~\cite{Adams:1992bn}.}
\useful{Review of axion inflation models:~\cite{Daido:2017wwb}. Discussion of super-Planckian decay constants in string theory can be found in Ref.~\cite{Banks:1984tw}. Realisations within modified gravity frameworks that can rescue the model with sub-Planckian $f$ can be found in Refs.~\cite{Palti:2015xra, Bostan:2025vkt}.}
\variation{
\begin{itemize}[leftmargin=1em,  itemsep=1pt]
    \item \textit{Multi-Natural Inflation:} A single axion obtains its potential from multiple cosine terms, allowing for a flat potential suitable for inflation even with sub-Planckian decay constants~\cite{Czerny:2014wza,Czerny:2014xja}.
    \item \textit{Palatini Natural Inflation:} Embedding the model in Palatini gravity with non-minimal couplings to gravity or additional scalar fields can flatten the potential, allowing for successful inflation with sub-Planckian decay constants~\cite{Racioppi:2018zoy, Racioppi:2017spw, Bostan:2022swq, Bostan:2025vkt, Bostan:2025zdt}.
    \item \textit{Warm Natural Inflation:} Dissipative effects (thermal friction) allow for slow-roll on steeper potentials, rescuing the model for small $f$~\cite{Visinelli:2011jy}.
    \item \textit{Axion Alignment/N-flation:} Multiple axions can generate an effective super-Planckian decay constant from sub-Planckian fundamental scales~\cite{Kappl:2014lra, Dimopoulos:2005ac}.
\end{itemize}
}
\comments{The requirement of a super-Planckian decay constant $f$ presents a significant challenge for \ac{UV} completions, particularly in string theory where $f \lesssim M_{\mathrm{Pl}}$ is typical. This motivated the development of multi-axion models like N-flation and recent frameworks in Palatini $F(R,X)$ gravity that can achieve successful inflation with sub-Planckian $f$. For related single-axion variants with multiple cosine terms, see also the multi-natural inflation discussed in the \ac{ALP} Inflation (Sec.~\ref{sec:Alp_inflation}) and \ac{ALP} Miracle (Sec.~\ref{sec:Alp_miracle}).}

\subsubsection{N-flation (Assisted Inflation)\hfill \hierarchy\inflation}
\purpose{To achieve effective large-field inflation with multiple axion-like particles, each with sub-Planckian decay constants, circumventing the theoretical challenges of single-field super-Planckian excursions.}
\modelling{The model features $N \gg 1$ uncoupled, light axion-like fields (canonical \ac{WISP}s) with random initial conditions, each with a periodic potential $V_i(\phi_i) = \Lambda_i^4 [1 - \cos(\phi_i/f_i)]$. The collective dynamics of these $N$ fields drive inflation, with the combined effect mimicking a single field with an effective displacement $\phi_{\text{eff}} \sim \sqrt{N}\langle \phi \rangle$. The mass spectrum is approximately $m^2 \sim \Lambda^4/f^2$, and the fields interact only gravitationally.}
\paper{The concept of assisted inflation was introduced in Ref.~\cite{Liddle:1998jc}, while the specific N-flation realisation with axions was proposed in Ref.~\cite{Dimopoulos:2005ac} and further analysed in Ref.~\cite{Easther:2005zr}.}
\useful{Connection to the string axiverse:~\cite{Arvanitaki:2009fg}, analysis of density perturbations:~\cite{Mukhanov:1997fw}, lattice studies of preheating:~\cite{Battefeld:2008rd,DeCross:2016fdz}. Comparison with other multi-field models:~\cite{Achucarro:2022qrl}.}
\bounds{The model predicts a scalar spectral index and tensor-to-scalar ratio consistent with \emph{Planck}~\cite{Planck:2018vyg} constraints for viable parameter choices. The decay constants are constrained by the WGC to $f_i \lesssim M_{\text{Pl}}$, and the number of fields required is typically $N \sim 10^2$-$10^3$ for $f_i \sim 0.1 M_{\text{Pl}}$. Isocurvature perturbations are suppressed due to the large number of fields.}
\comments{N-flation provides a compelling \ac{UV}-completion of large-field inflation using only sub-Planckian axion decay constants. The model naturally connects to the string axiverse scenario, where hundreds of axions appear in compactifications. After inflation, the lightest axions in the spectrum become \ac{DM} candidates through misalignment (see \ref{sec:misalignment}), creating a direct link between inflationary dynamics and the contemporary \ac{WISP} landscape. The reheating process involves complex, multi-field dynamics and can produce gravitational waves.}

\subsubsection{Oscillons/I-balls \hfill\darkmatter}
\purpose{To form long-lived, localised and oscillating non-topological solitons from a real scalar field with attractive self-interactions, providing a novel mechanism for scalar field condensation and potential \ac{DM} substructure.}
\modelling{Localised, non-perturbative field configurations. It is useful to distinguish between:
\begin{itemize}[leftmargin=1em,  itemsep=1pt]
    \item \textit{Oscillons:} Formed from \textit{real} scalar fields with a potential flatter than quadratic (attractive self-interaction). They are metastable and long-lived due to the approximate conservation of an adiabatic invariant, though they remain strictly unstable due to the classical radiation of scalar waves.
    \item \textit{I-balls (or Q-balls):} Formed from \textit{complex} scalar fields with a global $U(1)$ symmetry. They can be absolutely stable if the energy per particle in the soliton is lower than the mass of the free particle, protected by the conserved Noether charge.
\end{itemize}}
\paper{Oscillons were identified as long-lived configurations in Refs.~\cite{Gleiser:1993pt, Copeland:1995fq}, while the concept of I-balls (or Q-balls) was developed in Ref.~\cite{Kasuya:2002zs}, with specific application to \ac{ALP} in Ref.~\cite{Kawasaki:2020jnw}.}
\useful{Review of oscillons in axion-like fields~\cite{Amin:2013ika,Alexeeva:2023rfi}. Connection to ultra-compact minihalos and gravitational waves~\cite{Antusch:2016con, Zhou:2013tsa}.}
\bounds{Constrained by non-observation of anomalous gravitational microlensing~\cite{Fairbairn:2019xog}, diffuse gamma-ray background from decays~\cite{Amin:2019ums} and effects on large-scale structure formation. Future gravitational wave observatories could probe the parameter space through merger signals.}
\comments{Oscillons are particularly relevant for axion-like particles and other scalar \ac{WISP}s with periodic or flattened potentials. They can significantly alter the thermal history after inflation, delay thermalisation and generate unique gravitational wave signatures. While Q-balls can serve as stable \ac{DM} candidates, oscillons are typically transient; however, they can dominate the energy density of the early Universe, delay reheating and leave signatures in the form of gravitational waves or primordial black holes.}

\subsubsection{Scalaron (Starobinsky \texorpdfstring{$\mathcal{R}^2$}{R2}) Inflation \hfill\hierarchy\inflation}
\label{subsec:R2}
\purpose{To provide a successful model of inflation driven by a purely gravitational degree of freedom, predicting a specific, favoured tensor-to-scalar ratio.}
\modelling{The inflaton (scalaron) is a propagating scalar degree of freedom arising from the $\mathcal{R} + \mathcal{R}^2/6M^2$ term in the modified gravitational action (Metric formalism). In the Einstein frame, this becomes a canonical scalar field with a plateau-like potential. The scalaron mass is $M \approx 3 \times 10^{13}$ GeV, fixed by the amplitude of scalar perturbations.}
\paper{The mechanism was originally proposed in Ref.~\cite{Starobinsky:1980te}.}
\useful{Review of $f(R)$ gravity and the scalaron:~\cite{Sotiriou:2008rp,DeFelice:2010aj}. Connection to Higgs inflation:~\cite{Bezrukov:2007ep} and No-Scale Supergravity:~\cite{Ellis:2013xoa}.}
\bounds{The model predicts a scalar spectral index  and a tensor-to-scalar ratio values, which is consistent with \emph{Planck}/BICEP/\emph{Keck} data~\cite{Planck:2018vyg,BICEP:2021xfz}.}
\comments{The scalaron couples to the trace of the stress-energy tensor. Note that in the \textbf{Palatini formalism}, a pure $\mathcal{R}^2$ term does \textit{not} propagate an extra scalar degree of freedom. However, in the presence of other matter fields (e.g., axions), the Palatini $\mathcal{R}^2$ term generates non-minimal interactions that can ``flatten'' the matter potential, enabling inflation with sub-Planckian decay constants.}

\subsubsection{Symmetron \hfill\darkenergy \darkmatter \inflation}
\label{sec:Symmetron}
\purpose{Proposed as a potential \ac{DE} candidate. Its screening mechanism was introduced as a novel way of circumventing Solar System constraints on fifth forces expected from scalar fields in modified gravity theories.}
\modelling{The symmetron potential has a $\mathbb{Z}_2$ symmetry, which is broken in environments of low densities and, therefore, gives rise to a non-vanishing symmetron \ac{VEV}. However, in sufficiently dense environments, the symmetry is restored, leading to a vanishing \ac{VEV} and a suppression of the fifth force. This is a particular realisation of the Damour-Polyakov mechanism.}
\paper{The model was originally proposed in Refs.~\cite{Hinterbichler:2010es,Hinterbichler:2011ca}.}
\useful{Discussions of symmetron fifth forces as alternatives to particle \ac{DM} can be found in Refs.~\cite{Burrage:2016yjm,OHare:2018ayv,Burrage:2018zuj}. A radiatively stable symmetron model can be found in Ref.~\cite{Burrage:2016xzz}. The symmetron has also inspired a new inflationary scenario~\cite{Dong:2013swa}.}
\bounds{All currently existing constraints on this model can be found in Refs.~\cite{Burrage:2017qrf,Fischer:2024eic}. Further constraints from neutrino propagation can be found in Ref.~\cite{deGiorgi:2025kyp}.}
\comments{There also exist generalisations of this model to higher power potentials (see \hyperref[sec:GenSymmetrons]{Generalized symmetrons}). Ref.~\cite{Udemba:2025csd} indicates that quantum corrections might weaken the symmetron fifth force.}

\subsubsection{Warm Inflation \hfill\hierarchy}
\purpose{To provide a consistent inflationary paradigm where the inflaton's energy is continuously dissipated into a thermal bath during inflation, avoiding the need for a separate reheating epoch and naturally connecting to the radiation-dominated era.}
\modelling{The inflaton (a scalar \ac{WISP}) is coupled to other fields (e.g., a thermal bath of light particles) such that its slow-roll is maintained by both Hubble friction and dissipative effects. This dissipation, characterised by a coefficient typically labelled by $\Upsilon$, generates radiation concurrently with inflation, keeping the Universe at a non-zero temperature $T > H$. The inflaton's \ac{WISP}-like nature, its light mass and weak couplings, is crucial for sustaining a consistent warm inflation regime without triggering large thermal corrections that would spoil the flatness of the potential.}
\paper{The foundational framework was established in~\cite{Berera:1995ie}. Key developments on the dynamics and model-building were provided in~\cite{Berera:1999ws, Hall:2004ab}.}
\useful{Comprehensive reviews:~\cite{Hall:2004ab, Bastero-Gil:2019gao}. Analysis of dissipation mechanisms and quantum corrections can be found in Ref.~\cite{Moss:2011qc}. Important work on the statistical mechanics of warm inflation and its observational imprints, including non-Gaussianity, can be found in~\cite{Bastero-Gil:2018yen}.}
\bounds{The primary constraints come from the amplitude and shape of the \ac{CMB} power spectrum~\cite{Planck:2018vyg}. The model predicts a slightly redder spectral tilt and enhanced non-Gaussianity compared to cold inflation, which can be tested with future \ac{CMB} observations~\cite{Hazumi:2019lys, SimonsObservatory:2018koc}. The requirement of a sustained thermal bath without overproducing relics (e.g.\ gravitinos or moduli) also places bounds on the dissipative coefficient and the temperature during inflation.}
\comments{Warm inflation represents a distinct dynamical paradigm where the inflaton acts as a \ac{WISP} interacting with a thermal environment. It elegantly solves the ``eta problem" for certain potentials by leveraging dissipative effects. The model is particularly well-motivated in supersymmetric setups where the flatness of the inflaton potential is protected, and the necessary couplings for dissipation arise naturally. Furthermore, recent developments show that warm inflation can be successfully realised in extremely minimal models, such as a pseudo-scalar inflaton featuring an axion-like coupling to \ac{QCD} gluons and monomial potential~\cite{Berghaus:2025dqi}. In these setups, the shift symmetry associated with the \(aG\tilde{G}\) coupling protects the flat potential from the dangerous thermal mass corrections that typically plague warm inflation models. Phenomenologically, this is highly compelling as it implies the inflaton could be directly probed by standard axion search experiments.}

\subsubsection{Wallion \hfill\darkenergy\darkmatter}
\purpose{
Fields with a bounded field space exhibit a mirage cutoff, which yields a flat and radiatively stable potential away from the boundaries, motivating (ultra)light scalar dark matter.
}
\modelling{
The wallion is a real scalar field $\phi$ constrained to a field range of $|\phi| \lesssim \bar\phi$. 
The exponential suppression of the mass, $m_\phi \sim \Lambda\, \exp \left(-\bar\phi^2/(2\Lambda^2) \right)$,
arises when the boundary separation $\bar\phi$ exceeds the intrinsic boundary width $\Lambda$. 
Field-space boundaries may be generated dynamically via instantons, requiring a quadratic coupling of $\phi$ to the gauge kinetic term of a confining gauge group with a sign opposite to that of the gauge kinetic term.}
\paper{The concept of field-space boundaries and the associated mirage cutoff was introduced in Ref.~\cite{Cheung:2024wme}. 
}
\useful{
The wallion as a dark matter candidate was introduced in Ref.~\cite{Becker:2025pgb}.
Field-space boundaries have been applied to dark energy, where they allow for a technically natural quintessence potential fitting current observational data~\cite{Borghetto:2025jrk}.}
\bounds{
As a light scalar, the wallion can generate isocurvature and black hole superradiance. 
The instanton-generated wallion can result in quadratic wallion--\ac{SM} interactions. 
Depending on the wallion mass, constraints arise from equivalence principle tests, searches for time variation of fundamental constants, structure formation, and limits on $N_\text{eff}$. }
\comments{
Wallion potentials yield a late-time energy density independent of the initial displacement when the latter is sufficiently large, suppressing isocurvature perturbations relative to the axion case.
Periodic axion and wallion potentials belong to the same class of \ac{EFT}s subject to a mirage cutoff. The potentials differ in that a cosine axion potential yields alternating-sign coefficients, whereas the boundary-induced wallion potential produces coefficients of definite sign.
The wallion is a newly proposed \ac{WISP} with a single dedicated paper to date. 
Further wallion realizations and their phenomenological implications remain to be explored.}

\subsection{Composite Models}
\subsubsection{Composite Dark Matter \hfill\darkmatter}
\purpose{The purpose is to extend QCD-like theories confining at low energies, where the lowest mass state is naturally a stable \ac{DM} candidate.}
\modelling{$\mathrm{SU}(N_c)$ theories confine at low energies, where, depending on the theory, the \ac{DM} candidate can be either a meson or a baryon.}
\paper{The original idea in the context of technicolour was presented in Ref.~\cite{Nussinov:1985xr}. Different types of \ac{DM} models include Mesons~\cite{Kilic:2009mi,Bai:2010qg,Alves:2009nf}, Baryon-like \ac{DM}~\cite{Nussinov:1985xr,Chivukula:1989qb,Barr:1990ca}, and Glueball-like~\cite{Faraggi:2000pv} states.}
\useful{A review on strongly-coupled composite \ac{DM} models can be found in Ref.~\cite{Kribs:2016cew}.}
\bounds{Some constraints include model dependent \ac{DM} detection~\cite{Cappiello:2020lbk,Monteiro:2020wcb}, collider searches~\cite{Strassler:2006im, Han:2007ae} and Dark Showers~\cite{Baumgart:2009tn}.}
\comments{Strongly coupled theories have self-interactions, interesting for galactic structure-formation anomalies~\cite{Cline:2013zca}.}

\subsubsection{Nuclear/Molecular Dark Matter \hfill\darkmatter}
\purpose{The model investigates the possibility of \ac{DM} being made of a multi-constituent bound state. 
}
\modelling{These models typically introduce fermionic or 
bosonic dark constituents charged under a confining gauge group or a dark \(U(1)\), thereby producing dark ``nuclei'' or ``atoms'' with large occupation numbers \(N\). The resulting hierarchy between constituent mass, binding momentum, and composite size implies non-trivial momentum-dependent form factors in scattering processes.} 

\paper{References~\cite{Hardy:2015boa,Detmold:2014qqa,Cline:2021itd} explain the nature of a confining dark sector, leading to composite states such as dark baryons, dark nuclei, and dark atoms. They study the underlying gauge dynamics, binding energies, and dark nuclear reactions, including early-Universe dark nucleosynthesis, to determine relic abundances. Self-interactions, mass scales, and couplings relevant for phenomenology beyond elementary \ac{DM} candidates are also analysed.}
\useful{A discussion on composite inelastic \ac{DM} models where dark-sector bound states exhibit low-lying excited and generalisation of this idea to loosely bound \ac{DM} is described in Refs.~{\cite{SpierMoreiraAlves:2010err,Acevedo:2024lyr}}.}
\bounds{Current bounds on composite \ac{DM} restrict effective nucleon scattering cross sections and spatial extent by combining direct detection exclusions with phenomenology of propagation through matter. Extended composites that disassemble in terrestrial media exhibit unique multi-scatter and cascade signatures that probe parameter space beyond point-like limits. These bounds imply that viable composite \ac{DM} models must have suppressed interactions, strong form-factor effects, or occupy constrained regions of mass and cross-section~{\cite{Boukhtouchen:2025vvg}}.}

\section{Spin-1/2}
\subsubsection{Dark Photini \hfill \darkmatter\stringtheory}
\label{sec:dark-photini}
\purpose{In supersymmetric models, extra $U(1)$ gauge bosons are accompanied by their fermionic superpartners. }
\modelling{There exists a natural supersymmetric version of the kinetic mixing term. Even if the $U(1)$ gauge symmetry is unbroken, supersymmetry breaking gives masses to the dark photini which can then interact with the Standard Model via kinetic/mass mixing with the photino.}
\paper{Early discussions of these models appear in Refs.~\cite{Ibarra:2008kn,Arvanitaki:2009hb}. Motivation from string theory is given in Ref.~\cite{Coudarchet:2025dfd}.
}
\bounds{Reference~\cite{Ibarra:2008kn} discusses dark matter and constraints arising from it and Ref.~~\cite{Arvanitaki:2009hb} investigates \ac{LHC} phenomenology.} 
\comments{String models suggest that dark photini may be lighter than the other \ac{SUSY} particles~\cite{Coudarchet:2025dfd}, making them a candidate for dark matter and more accessible in phenomenology.}

\subsubsection{Kaluza-Klein Neutrinos \hfill\genesis\flavour\neutrinos\stringtheory}
\label{sec:KK-sterile-neutrinos}
\purpose{Generate neutrino masses and/or constrain extra dimensions by studying the impact on neutrino phenomenology.}
\modelling{In the presence of extra dimensions, if the \ac{SM} is localised on a 3D-brane, any fermion propagating in the extra-dimensional bulk is a \ac{SM} gauge singlet, and thus qualifies as a sterile neutrino. In the simplest scenario, it amounts to a free bulk fermion in flat space. 
The tower of sterile neutrinos from the extra dimension can impact neutrino phenomenology, such as mass splittings and thus the oscillation phenomenology. In the simplest scenario, the spectrum is determined entirely by the geometry of the extra dimension (e.g.\ flat vs warped).
The model can be made arbitrarily complicated by including Dirac or Majorana mass terms in the bulk or on the brane.}
\paper{The flat case has been first studied in Refs.~\cite{Dienes:1998sb,Arkani-Hamed:1998wuz,Dvali:1999cn,Lukas:2000rg}, while warped geometries were examined in Refs.~\cite{Grossman:1999ra,Huber:2003sf,Fong:2011xh}.}
\useful{An overview of the spectra in different setups in 5D flat space can be found e.g.\ in Ref.~\cite{deGiorgi:2025xgp}.}
\bounds{Constraints on flat extra-dimensional scenarios from neutrino oscillation in the Dirac case can be found e.g.\ in Ref.~\cite{Elacmaz:2025ihm}, while constraints for Majorana scenarios can be found e.g.\ in Ref.~\cite{deGiorgi:2025xgp}.}
\comments{The model is effectively a generalisation of the \hyperref[sec:sterile-neutrinos]{sterile neutrino} scenario, where an infinite tower of sterile neutrinos is included. Recently, it has been pointed out that conjectures of the Swampland programme allow for a mesoscopic ``Dark Dimension" scenario~\cite{Montero:2022prj}, which provides a modern theoretical motivation for a tower of KK neutrinos at the meV scale, tied to the vacuum energy density.}

\subsubsection{Sterile Neutrinos \hfill\genesis\darkmatter\neutrinos}
\label{sec:sterile-neutrinos}
\purpose{The purpose is to give mass to the \ac{SM} neutrinos. Possibly, they can contribute to explaining the \ac{DM} abundance {and the baryon asymmetry}.}
\modelling{The model includes new fermions sterile under the \ac{SM} gauge group, typically dubbed as ``right-handed" or ``sterile" neutrinos. The landscape of sterile neutrino models is very vast, both in mass scales and Lagrangians.}
\paper{
Seesaw constructions were first proposed in Refs.~\cite{Minkowski:1977sc,Gell-Mann:1979vob,Yanagida:1979as,Mohapatra:1979ia}. 
Leptogenesis was originally proposed in Ref.~\cite{Fukugita:1986hr} via the decay of thermally produced right-handed neutrinos, with the generated lepton asymmetry converted into a baryon asymmetry through \ac{EW} sphaleron processes. 
It can also be realized if right-handed neutrinos are produced non-thermally~\cite{Lazarides:1991wu}. 
The viable mass range can be lowered, and the required reheating temperature reduced, in the presence of nearly degenerate right-handed neutrinos (resonant leptogenesis)~\cite{Pilaftsis:1997dr,Buchmuller:1997yu}. 
Alternative mechanisms include leptogenesis via flavor oscillations of right-handed neutrinos (ARS mechanism)~\cite{Akhmedov:1998qx} or via left-handed neutrino oscillations during reheating~\cite{Eijima:2019hey}, both of which allow for very low-scale leptogenesis. 
The framework known as the $\nu$MSM, which attempts to explain both \ac{DM} and baryogenesis by introducing three right-handed neutrinos, was proposed in Refs.~\cite{Asaka:2005pn,Asaka:2005an}. 
See also Refs.~\cite{Ghiglieri:2020ulj,Canetti:2012kh} for discussions of dark matter and baryon asymmetry in the $\nu$MSM, and Ref.~\cite{Yin:2024trc} for asymmetric sterile-neutrino dark matter.}

\useful{An overview of the topic can be found in Refs.~\cite{Drewes:2013gca,Drewes:2016upu,Boyarsky:2018tvu}.}
\bounds{General constraints from global fits on $eV$ sterile neutrinos can be found in Refs.~\cite{Dentler:2018sju,Diaz:2019fwt,Boser:2019rta,Hagstotz:2020ukm,Hardin:2022muu}. For the sterile \ac{DM} with a mass in the keV range~\cite{Blennow:2023mqx}, there exist strong constraints from $X$-ray and $\gamma$-ray observations, as well as from small-scale structure formation. More massive candidates have been considered astrophysical sites~\cite{Albertus:2015xra} showing a potentially large impact on supernovae rates~\cite{Rembiasz:2018lok}. In addition, the Tremaine--Gunn bound~\cite{Tremaine:1979we,Boyarsky:2008ju} provides a model-independent lower limit on the \ac{DM} mass. Further constraints on heavy sterile neutrinos from unitarity violation of the PMNS can be found in Ref.~\cite{Blennow:2023mqx}. 
See also {Ref.~\cite{Hernandez:2022ivz} for bounds coming from observable leptogenesis and Ref.~\cite{Drewes:2021nqr} for the limits for the ARS and resonant Leptogenesis mechanisms.} 
} 
\comments{Currently, at least two right-handed neutrinos are needed to explain neutrino oscillations. {The minimal thermal production via the Dodelson–Widrow mechanism~\cite{Dodelson:1993je} is in tension with the current limits~\cite{Boyarsky:2018tvu}. New mechanisms have been proposed for sterile neutrino \ac{DM}, such as production from a large lepton asymmetry~\cite{Shi:1998km}, inflaton decay~\cite{Bezrukov:2014nza,Moroi:2020has}. Some scenarios even allow \ac{DM} itself to be responsible for neutrino oscillations.}}


\section{Spin-1}
\subsubsection{B-L Dark Photon \hfill\neutrinos}
\purpose{To gauge the unique anomaly-free global symmetry of the \ac{SM} in order to explain the origin of neutrino masses.}
\modelling{The gauge group of the \ac{SM} is extended by $U(1)_{B-L}$, giving rise to a massive vector boson $Z'$ that couples to fermions with strength proportional to their $B-L$ charge. The $B-L$ boson couples directly to neutrinos, and anomaly cancellation requires three right-handed neutrinos.}
\paper{The foundational ideas of the model were placed in Refs.~\cite{Davidson:1978pm,Marshak:1979fm, Mohapatra:1980qe, Wetterich:1981bx}.}
\useful{Ref.~\cite{Bauer:2018onh} translates generic dark photon limits into $B-L$ coupling limits.}
\bounds{A collection of bounds can be found in Refs.~\cite{Heeck:2014zfa,AxionLimits}. Further constraints come from neutrino scatterings~\cite{Bilmis:2015lja}, FASER~\cite{FASER:2023tle}, binary pulsars~\cite{LopezNacir:2018epg} and LIGO--Virgo--KAGRA~\cite{LIGOScientific:2025ttj}. Constraints on light mediators from $N_\text{eff}$ for different realisations of the model can be found in Ref.~\cite{Ghosh:2024cxi}.}
\comments{A $U(1)_B$-only boson was proposed in~\cite{Carone:1994aa}, anomaly cancellation and possible \ac{DM} candidates were first discussed in~\cite{FileviezPerez:2010gw}.}

\subsubsection{Dark Photon EFT \hfill\none}
\purpose{Generic extension of the \ac{SM} to exploit the ``vectorial portal" to a dark sector.}
\modelling{The dark photon is coupled to the \ac{SM} via a mixing with the kinetic terms of the gauge fields; a mass term is typically introduced by hand.}
\paper{It is difficult to point to a few papers that define the field. Some of them can be found in Refs.~\cite{Fayet:1980rr,Okun:1982xi,Georgi:1983sy,Holdom:1985ag}.}
\useful{For an overview of the topic, see e.g.\ Refs.~\cite{Fabbrichesi:2020wbt,Caputo:2021eaa,Caputo:2026pdw}.}
\bounds{A collection of existing constraints is given in Ref.~\cite{AxionLimits}; superradiance constraints are in Ref.~\cite{Unal:2020jiy}; the most recent ALPS-II limits are in Ref.~\cite{ALPSII:2025eri}. One can use the available DarkCast public code to extract bounds from terrestrial experimental searches, including vector and axial couplings, see Refs.~\cite{Ilten:2018crw,Baruch:2022esd}.}
\comments{In some parts of the parameter space, the bounds on the model can be made significantly stronger when a full \ac{UV} model is considered.
The model can accommodate \ac{DM} abundance.
The original misalignment mechanism was discussed in Ref.~\cite{Nelson:2011sf}, but it was later pointed out that the initial misalignment amplitude is diluted during inflation~\cite{Arias:2012az}. To alleviate this issue, non-minimal couplings have been introduced, although such extensions typically lead to a ghost instability. This instability can be resolved either by introducing an inflaton-dependent kinetic function together with a curvaton~\cite{Kitajima:2023fun}, or by embedding the $U(1)$ gauge symmetry into a larger $SU(N)$ gauge structure~\cite{Fujita:2023axo}. }

\subsubsection{Geometric $Z'$ \hfill \darkmatter}
\purpose{To create a massive vector \ac{WISP} candidate of purely geometric origin emerging from \ac{MAG}, potentially viable as \ac{DM}.}
\modelling{\ac{MAG} with independent metric and affine connection, with action as in Ref.~\cite{Demir:2020brg}; after decomposing the connection (torsion-free) the theory dynamically reduces to \ac{GR} plus a massive vector (geometric Proca) coupled only to \ac{SM} fermions.}
\paper{The model was proposed in Refs.~\cite{Demir:2020brg,Demir:2022wpp}.}
\useful{Refs.~\cite{Demir:2022wpp,Ghorani:2023hkm,Ghorani:2024ufk} provide black-hole solutions and phenomenology with the geometric vector field.}
\bounds{Bounds are studied in Ref.~\cite{Demir:2020brg}, including cosmological-lifetime and kinematic consistency constraints leading to the allowed MeV-scale mass window.}
\comments{A distinctive feature of the new vector is that it couples exclusively to fermions (no couplings to scalars/gauge bosons), yielding extremely feeble direct-detection prospects while remaining theoretically self-consistent in \ac{MAG}~\cite{Demir:2020brg}.}


\subsubsection{Leptophilic $Z'$ \hfill\darkmatter}
\purpose{Building a $Z'$ boson with anomaly-free U(1)’ models with exclusively leptonic gauge charges.}
\paper{The model was proposed in Ref.~\cite{He:1991qd}}
\bounds{Many bounds have been extracted using various methods, such as neutrino oscillation experiments, \ac{DM} direct detection, solar neutrino, and nuclear reactors~\cite{AxionLimits}. Furthermore, further constraints from Majoron-type physics apply (see e.g. Ref.~\cite{Fiorillo:2022cdq}).}
\comments{The model predicts an extra \ac{BSM} vector mediator that potentially mediates neutrino interactions with the ordinary \ac{SM} particles. The mediator is also regarded as a dark photon-like particle that can interact with the $Z$ boson with kinetic mixing or mass mixing terms.}

\subsubsection{Massless Dark Photons with Dipole Interactions \hfill \stringtheory}
\purpose{Model interactions of extra $U(1)$ gauge bosons which are often massless, and thus typically unobservable.}
\modelling{If SM matter is uncharged under the new $U(1)$, the leading kinetic mixing term is unobservable. Nevertheless, it can still interact via dimension-6 dipole interactions with SM fermions. In the presence of such interactions, kinetic mixing also becomes observable and leads to additional effective electric and magnetic dipole moments with respect to the ordinary photon.}
\paper{The model was introduced in Ref.~\cite{Dobrescu:2004wz} and identified as a generic string prediction in Ref.~\cite{Coudarchet:2025dfd}.}
\useful{The original Ref.~\cite{Dobrescu:2004wz} already includes phenomenological tests. Spin-dependent forces were discussed in~\cite{Dobrescu:2006au} and some first astrophysical constraints in~\cite{Camalich:2020wac}. More generally, the phenomenology was reviewed and updated in Refs.~\cite{Fabbrichesi:2020wbt,Coudarchet:2025dfd}. \cite{Coudarchet:2025dfd} estimated the size of the dimension 6 dipole operators in string models and discussed the interplay of the dipole moments and kinetic mixing. }
\bounds{Some recent overview of limits can be found in Refs.~\cite{Fabbrichesi:2020wbt,Coudarchet:2025dfd}.} 
\comments{Massless dark photons cannot be dark matter. But the \hyperref[sec:dark-photini]{dark photini} that accompany them in supersymmetric theories may play this role for specific parameter values~\cite{Ibarra:2008kn,Arvanitaki:2009hb}.}


\subsubsection{Millicharged Particles Mediator \hfill\darkmatter}
\purpose{\ac{mCPs} arise from models where the visible and dark sectors are kinetically mixed or from models where these particles are directly charged under the $U(1)_D$ in the \ac{UV} theory. They are also widely predicted in string theory compactifications and Grand Unification theories.}
\modelling{Millicharged \ac{DM} arises naturally in theories containing hidden sectors charged under an additional gauge symmetry \(U(1)_D\) with a corresponding dark photon \(A'_\mu\). The key mechanism that generates a small effective electric charge is the gauge- and Lorentz-invariant kinetic mixing term \(\frac{\epsilon}{2} F^{\mu\nu} F'_{\mu\nu}\), which appears radiatively in many extensions of the \ac{SM}. Strictly speaking, the millicharged particle itself is usually a fermion (spin-1/2) or scalar (spin-0), while the mediator is the spin-1 state. We focus here on the latter. If a dark-sector field \(\chi\) carries charge \(g_D\) under \(U(1)_D\), the mixing  induces an effective visible-sector charge \(q_{\mathrm{eff}} = \epsilon\, g_D\). This structure arises generically in supergravity models, string compactifications with multiple Abelian factors and heterotic constructions where kinetic mixing is ubiquitous. Similar millicharged states emerge in Hidden Valley frameworks, gauge-seesaw models and mirror-matter scenarios that replicate the \ac{SM} gauge content. For sufficiently small \(\epsilon \lesssim 10^{-10}\), \ac{mCPs} remain consistent with cosmological, astrophysical and laboratory constraints while still allowing phenomenologically relevant interactions in the early Universe and in small-scale structure formation.}
\paper{The first realisation of the model was in Ref.~\cite{Holdom:1985ag}. A Stueckelberg variant is developed in Ref.~\cite{Feldman:2007wj}.}
\useful{An early string-theory embedding can be found in Ref.~\cite{Wen:1985qj}. Tests of stringy millicharged particles in the lab were given in Ref.~\cite{Cicoli:2011yh}, while Ref.~\cite{Kouvaris:2014rja} discusses the effects on neutron-star spin-down.}
\bounds{Recent constraints on mCPs span many orders of magnitude in mass and charge, reflecting the diversity of observational and experimental probes. 
Ultralight candidates are tightly constrained by precision electromagnetic measurements, including timing of radio waves, SNe and geomagnetic/magnetometer observations, which typically bound charge-to-mass combinations rather than a single mass-independent value~\cite{Caputo:2019tms, Fiorillo:2024upk,Arza:2025cou, Fiorillo:2024upk}.  
For sub-MeV masses, laboratory sensitivity is being actively developed with novel techniques such as ion-trap searches~\cite{Budker:2021quh}.  
In the MeV–GeV window, a variety of accelerator and laboratory searches constrain millicharges, with the precise exclusions depending on the production mechanism.  
At GeV and higher masses, plasma and astrophysical observations such as the Bullet Cluster provide additional stringent constraints~\cite{Medvedev_2025}.}
\comments{This model shares similarities with those \ac{DM} candidates under the form of strange matter droplets (strangelets) as they may be slightly charged~\cite{AngelesPerez-Garcia:2013fpo,DiClemente:2024lzi}.}

\section{Spin-3/2}

\subsubsection{Gravitino \hfill\darkmatter\hierarchy}
\purpose{The supersymmetric extension of the \ac{SM} and gravity predicts the existence of the gravitino, the superpartner of the graviton, which can serve as a viable \ac{DM} candidate if it is the \ac{LSP} and if the R-parity is imposed. In a gauge-mediated supersymmetry breaking scenario~\cite{Dine:1993yw}, the gravitino \ac{LSP} is a natural prediction.}
\modelling{The relic abundance of gravitino \ac{DM} can arise from both thermal production in the early Universe and non-thermal production via decays of heavier particles. The exact abundance depends sensitively on the reheating temperature and the new particle mass spectrum.}
\paper{Early works on gravitinos as \ac{DM} are Refs.~\cite{Pagels:1981ke,Weinberg:1982zq,Ellis:1984eq}}
\useful{A comprehensive derivation of gravitino production and decay in cosmology with precise abundances and implications for \ac{BBN} and \ac{SUSY} can be found in Ref.~\cite{Moroi:1995fs,Eberl:2024pxr}.}
\bounds{Constraints arise from \ac{BBN} and small-scale structure formation, as well as from limits on late-time energy injection by the decay of next-to-lightest supersymmetric particles (NLSP). Accelerator experiment searching for long-lived NLSP is also relevant.}
\comments{If $R$-parity is slightly violated and the gravitino constitutes decaying \ac{DM}, it can be searched for through its decay~\cite{Takayama:2000uz,Buchmuller:2007ui,Ibarra:2007wg}.}

\section{Spin-2}
\subsubsection{Dark Graviton EFT \hfill \none}
\label{sec:graviton-EFT}
\purpose{Extension of General Relativity to include a new massive spin-2 degree of freedom.}
\modelling{A unique ghost-free potential (the dRGT potential) is constructed for the massive graviton, which requires a second metric (fixed in the original dRGT construction and fully dynamical in bigravity).}
\paper{Papers that fundamentally contributed in building the theory:~\cite{deRham:2010ik,deRham:2010kj,Hassan:2011zd,Hassan:2011vm,deRham:2013tfa}.}
\useful{A general review can be found in Ref.~\cite{Hinterbichler:2011tt}, while applications to \ac{DM} can be found in Refs.~\cite{Aoki:2016zgp,Babichev:2016bxi,Babichev:2016bxi}.}
\bounds{A compilation of constraints can be found in Ref.~\cite{Cembranos:2017vgi,Garcia-Cely:2025ula}. Constraints from LIGO--Virgo--KAGRA on \ac{DM} scenarios can be found in Ref.~\cite{LIGOScientific:2025ttj}, superradiance constraints are in Ref.~\cite{Unal:2020jiy}, pulsar-timing constraints are in Refs.~\cite{Armaleo:2019gil,Armaleo:2020yml}, while Ref.~\cite{Garcia-Cely:2025ula} obtains constraints from stellar emission.}
\comments{The model suffers from breaking at a (possibly) low-energy scale of perturbative unitarity when including the scattering of massive gravitons. The model also has a potentially low strong coupling scale. A chameleon variant exists that can evade fine-tuning and strong coupling problems~\cite{DeFelice:2017oym}; some of these difficulties can also be overcome at the expense of Lorentz invariance, see Refs.~\cite{Dubovsky:2004sg,DeFelice:2020eju}. Constraints on the \ac{EFT} scale from positivity bounds can be found in Refs.~\cite{Bellazzini:2023nqj,Dong:2025dpy}.}

\subsubsection{KK Gravitons \hfill\stringtheory}
\purpose{Extra-dimensional models can be used in attempts to unify forces and fields and explain hierarchies of scales or Lagrangian parameters.}
\modelling{In the minimal version, General Relativity is extended to extra dimensions. The massive gravitons arise as \ac{KK}-states of the metric. In this entry, we do not refer to any specific model or region of the parameter space, but rather to the generic feature of extra-dimensional constructions.}
\paper{The original papers, which proposed extra dimensions in the context of gravity-electromagnetism unification in $5D$, can be found in Refs.~\cite{Kaluza:1921tu,Klein:1926tv}.}
\useful{Details about unitarity the theory can be found in Refs.~\cite{Chivukula:2020hvi,deGiorgi:2020qlg}.}
\bounds{A general collection of bounds on a single massive graviton can be found in Ref.~\cite{Cembranos:2017vgi}, and thus they need appropriate rescaling.}
\comments{Specific realisations of the model are the \hyperref[sec:LED]{Large Extra Dimensions}~(LED or ADD) and the \hyperref[sec:Warped_Extra]{Randall-Sundrum} models. When more than one graviton is on-shell in a scattering process, resummation of multiple intermediate \ac{KK}-modes is necessary to raise the effective cutoff of the theory. See the latter section for further details.}

\subsubsection{Large extra dimensions \hfill \hierarchy\stringtheory}
\label{sec:LED}
\purpose{Provide a geometric solution to the hierarchy problem by lowering the fundamental Planck scale via large extra dimensions.}
\modelling{Gravity propagates in $(4+d)$ flat dimensions compactified at large radii, while \ac{SM} fields are localised on a brane. Gravity becomes weak because it is diluted in the extra-dimensional volume, and the four-dimensional theory contains a dense tower of light (e.g.~$\sim$eV) \ac{KK} gravitons.}
\paper{Originally proposed in Refs.~\cite{Antoniadis:1990ew,Arkani-Hamed:1998jmv,Antoniadis:1998ig}, from whose authors' surnames it is sometimes referred to by ADD in the literature.}
\useful{A thorough pedagogical discussion can be found in Ref.~\cite{Giudice:1998ck}. Feynman rules were obtained in Refs.~\cite{Han:1998sg,Giudice:1998ck}.}
\bounds{The strongest constraints come from fifth-force tests, such as torsion balance~\cite{Kapner:2006si} and astrophysics~\cite{Hannestad:2003yd,Hardy:2025ajb}. Complementary but less constraining bounds come from colliders~\cite{ATLAS:2021kxv}. 
} 
\comments{Given natural parameter choices, the case $d=1$ is excluded by sub-millimetre gravity tests, while $d\geq 2$ remains viable. A different part of the parameter space can be viable if curvature is included, e.g. in the \hyperref[sec:Warped_Extra]{Randall-Sundrum model}. See such a section for comments about the cutoff of the theory.}

\subsubsection{Randall-Sundrum Warped Extra Dimensions \hfill \hierarchy\stringtheory}
\label{sec:Warped_Extra}
\purpose{Provide a geometric solution to the hierarchy problem via the existence of a warped extra dimension.}
\modelling{The model includes a warped extra dimension equipped with an orbifold symmetry. The \ac{SM} sits on a 3-brane fixed point of the orbifold symmetry. Gravity propagates in a slice of five-dimensional Anti--de Sitter space, generating a localised massless graviton and a discrete tower of heavy ($\sim$TeV) \ac{KK} gravitons with enhanced couplings to matter.}
\paper{The model was proposed originally in Refs.~\cite{Randall:1999ee,Randall:1999vf}.}
\useful{Some useful references include comments about modulus stabilisation,~\cite{Goldberger:1999uk} and phenomenology of spin-2 resonances and couplings~\cite{Davoudiasl:1999jd}.}
\bounds{The main constraints come from collider searches looking for heavy resonances, e.g. Ref.~\cite{ATLAS:2021uiz}.}
\comments{The effective cutoff scale of the theory is larger than that of the \hyperref[sec:graviton-EFT]{Dark Graviton EFT}. For this to happen, all the tower of \ac{KK}-states must be included when computing observables. This has been shown explicitly, e.g. both in graviton-graviton~\cite{Chivukula:2023qrt,Chivukula:2024vgg} and graviton-matter~\cite{deGiorgi:2020qlg,Chivukula:2023sua,deGiorgi:2023mdy} scattering.}

\subsubsection{Lorentz-Violating Massive Gravity \hfill \darkmatter}
\purpose{To construct a consistent \ac{EFT} of massive gravity without the Boulware-Deser ghost and vDVZ discontinuity.}
\modelling{The model extends General Relativity by adding Lorentz-violating mass terms for the graviton. This is often modelled as a spontaneous breaking of Lorentz invariance. The most general mass Lagrangian (preserving 3D spatial rotational symmetry) has five mass parameters, $m_0$ to $m_4$.}
\paper{The first systematic works can be found in Refs.~\cite{Arkani-Hamed:2003pdi, Rubakov:2004eb, Dubovsky:2004sg}.}
\useful{A review can be found in Ref.~\cite{Rubakov:2008nh}, while the \ac{DM} scenario is discussed in Ref.~\cite{Dubovsky:2004ud}.}
\comments{The theory is healthy only for specific fine-tunings of the mass parameters. These tunings can be protected by an unbroken residual gauge symmetry. In combination with the dGRT potential, the Minimal Theory of Bigravity breaks 4D diffeomorphism invariance to propagate only four tensor degrees of freedom~\cite{DeFelice:2020eju}.}


\section*{Acknowledgements}
\label{sec:acknowledgments}
This article/publication is based upon work from COST Action \Cwisp CA21106, supported by COST (European Cooperation in Science and Technology). The authors thank the members of the Action, especially its chair, A. Mirizzi, for the stimulating environment from which this encyclopedia arises.

 C.A.\ and M.A.P.G.\ are supported by Junta de Castilla y León SA101P24, SA091P24, MICIU project PID2022-137887NB-I00, Gravitational Wave Network (REDONGRA) Strategic Network (RED2024-153735-E) from Agencia Estatal de Investigación del MICIU (MICIU/AEI/10.13039/501100011033).
This research was funded in whole or in part by the Austrian Science Fund (FWF) [10.55776/PAT8564023]. F.U.\ acknowledges support from the European Structural and Investment Funds and the Czech Ministry of Education, Youth and Sports (project No.\ FORTE---CZ.02.01.01/00/22\_008/0004632). 
The work of M.F.Z.\ is supported by the Spanish MIU through the National Program FPU (grant number FPU22/\hspace{0pt}03625). 
M.F.Z.\ and L.M.\ acknowledge support from the Spanish Research Agency (Agencia Estatal de Investigaci\'on) through the grant IFT Centro de Excelencia Severo Ochoa No CEX2020-001007-S and by the grant PID2022-137127NB-I00 funded by MCIN/\hspace{0pt}AEI/\hspace{0pt}10.13039/501100011033.
F.C.D. is supported by STFC consolidated grant ST/X003167/1. 
W.Y.\ is supported by JSPS KAKENHI Grant Nos.\  22K14029, 22H01215 and Selective Research Fund from Tokyo Metropolitan University.

For open access purposes, the authors have applied a CC BY public copyright license to any author-accepted manuscript version arising from this submission. 

\section*{Contributors}
The authors are grateful to several colleagues for useful comments which improved the document: M.~Becker, A.~Cheek, M.~Drewes, E.~Fernández-Martínez, D.~Fiorillo, I.~D.~Gialamas, J.~Jaeckel, A.~Lella, G.~Lucente, I.~Martínez~Soler, A.~Mirizzi, R.~Natale, P.~Paradisi, M.~Putti, P.~Quílez, M.~Reig, P.~Schwaller, G.~Servant, Y.~Soreq, C.~Ünal, E.~Vitagliano.


\section*{List of Acronyms}
\label{sec:Acronyms}

\begin{acronym}
    \acro{ALP}{Axion-Like Particle}
    \acro{BBN}{Big Bang Nucleosynthesis}
    \acro{BSM}{Beyond the Standard Model}
    \acro{CMB}{Cosmic Microwave Background}
    \acro{DE}{Dark Energy}
    \acro{DGP}{Dvali-Gabadadze-Porrati}
    \acro{DM}{Dark matter}
    \acro{DW}{Domain wall}
    \acro{EFT}{Effective Field Theory} \acro{EDMs}{Electric Dipole Moments}
    \acro{EW}{Electroweak}
    \acro{FCNCs}{Flavour Changing Neutral Currents}
    \acro{FN}{Frogatt-Nielsen}
    \acro{GR}{General Relativity}
    \acro{GUTs}{Grand Unification Theories}
    \acro{GW}{Gravitational Wave}
    \acro{HNL}{Heavy Neutral Lepton}
    \acro{KK}{Kaluza-Klein}
    \acro{LHC}{Large Hadron Collider}
    \acro{LSP}{Lightest Supersymmetric Particle}
    \acro{MAG}{Metric-Affine Gravity}
    \acro{mCPs}{Millicharged Particles}
    \acro{MFV}{Minimal Flavour Violation}
    \acro{MSSM}{Minimal Supersymmetric Standard Model}
    \acro{NGB}{Nambu-Goldstone boson}
    \acro{pNGB}{pseudo-Nambu-Goldstone Boson}
    \acro{PQ}{Peccei-Quinn}
    \acro{PTA}{Pulsar Timing Array}
    \acro{QCD}{Quantum Chromodynamics}
    \acro{RGE}{Renormalization Group Equations}
    \acro{SSB}{Spontaneous Symmetry Breaking}
    \acro{SM}{Standard Model}
    \acro{SUGRA}{Supergravity}
    \acro{SUSY}{Supersymmetry}
    \acro{UV}{Ultraviolet}
    \acro{VEV}{Vacuum Expectation Value}
    \acro{WISP}{Weakly Interacting Slim Particle}
    \acro{2HDM}{Two-Higgs Doublet Model}
\end{acronym}

\footnotesize

\bibliographystyle{BiblioStyle}
\bibliography{Bibliography}

\end{document}